\documentclass[aps,prl,showpacs,twocolumn,amsmath]{revtex4}
\usepackage{graphicx}
\pdfoutput=1
\newcommand{\bq}{\begin{equation}}
\newcommand{\eq}{\end{equation}}
\newcommand{\bn}{\begin{eqnarray}}
\newcommand{\en}{\end{eqnarray}}
\newcommand{\bsub}{\begin{subequations}}
\newcommand{\esub}{\end{subequations}}
\newcommand{\s}{\sigma}

\newcommand{\om}{\omega}
\newcommand{\rem}[1]{}

\begin{document}
\title{Single molecule detection of nanomechanical motion}
\author{Vadim Puller$^{1}$}
\author{Brahim Lounis$^2$}
\author{Fabio Pistolesi$^{1}$}
\affiliation{
$^1$
Univ. Bordeaux, LOMA, UMR 5798, F-33400 Talence, France.\\
CNRS, LOMA, UMR 5798, F-33400 Talence, France.\\
$^2$
Univ Bordeaux, LP2N,  F-33400 Talence, France\\
Institut d'Optique and CNRS, UMR 5298, Talence, France
}

\begin{abstract}
We investigate theoretically how single-molecule spectroscopy techniques can be used to perform fast and high resolution displacement detection and manipulation of nanomechanical oscillators, such as singly clamped carbon nanotubes.
We analyze the possibility of real time displacement detection by the luminescence signal and of displacement fluctuations by the degree of second order coherence.
Estimates of the electro-mechanical coupling constant indicate that intriguing regimes of strong back-action between the two-level system of a molecule and the oscillator can be realized.
\end{abstract}
\date{\today}

 \maketitle

\paragraph{Introduction}

Perfecting displacement detectors is the central motivation driving the research
in nanomechanics \cite{Zant} and a necessary step towards quantum
manipulation of mechanical degrees of freedom.
The main strategy is to couple the mechanical oscillator to a well
controlled quantum system whose state can then be very accurately measured.
A non exhaustive list of such detection systems includes
single-electron transistors \cite{SET_sensing}, SQUIDs \cite{SQUIDs}, Qbits \cite{Cleland}, point contacts \cite{Lehnert}, optical \cite{Cavities},
and microwave cavities \cite{microwave}.
A wide range of possibilities is opened by the emergence of a new class
of detectors, where the displacement of the nano-oscillator modulates the
energy splitting of a two-level system (TLS) \cite{Arcizet,Rabl}, which is then
measured via optical resonance spectroscopy.
In Ref. \cite{Arcizet} the displacement ($x$) of the nano oscillator
is detected by exploiting a Zeeman split TLS (Nitrogen Vacancy center)
embedded on the tip of the oscillator moving in a strongly
inhomogeneous magnetic field.

\begin{figure}[tbp]
  \includegraphics[width=3in]{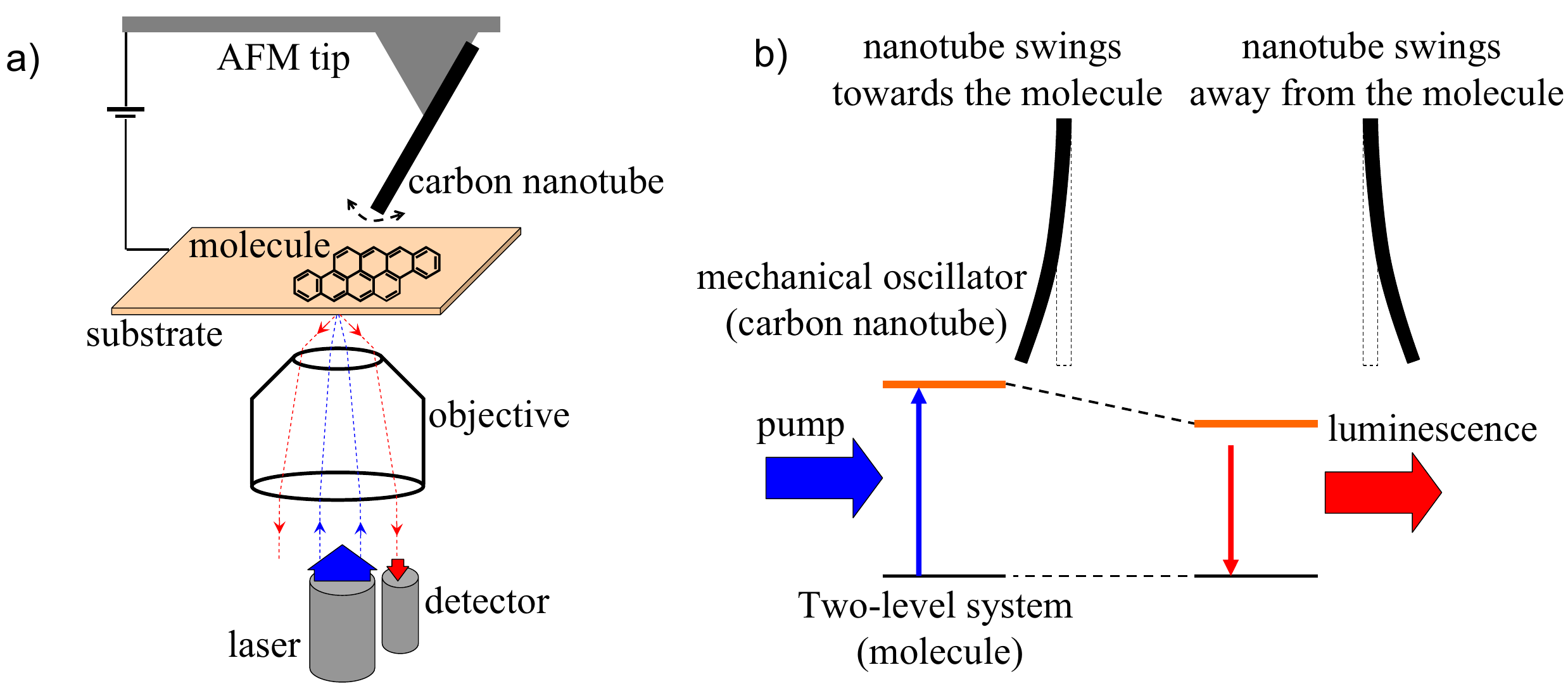}
  \caption{(Color online) Schematic representation of the proposed measurements setup: a) the geometry, b) modulation of molecular level splitting during the nanotube oscillations.}\label{fig:setup}
\end{figure}

In this Letter we propose using the single-molecule spectroscopy
(SMS) \cite{SMS_book,Tamarat,Stark_shift} to detect the
displacement of nano-oscillators,
such as carbon nanotubes suspended from the tip of an atomic
force microscope, see Fig. \ref{fig:setup}.
%
%%%%%%%%%%%%%%%%%%%%%%%%%%%
%                                                                        %
%                              FIGURE 1                             %
%                                                                        %
%%%%%%%%%%%%%%%%%%%%%%%%%%%
%
%
%%%%%%%%%%%%%%%%%%%%%%%%%%%%
%
At liquid helium temperatures, the zero-phonon lines of single-molecule
fluorescence excitation spectra are extremely narrow, since the dephasing
of optical transition dipole due to phonons vanishes.
For well chosen fluorophores/matrix systems \cite{SMS_book},
such as the dibenzo- anthanthrene DBATT embedded in a n-Hexadecane Shpol'skii matrix,
the zero-phonon has a spectral width limited by the lifetime of
the molecule excited state ($\sim 10- 20$ MHz) \cite{Stark_shift,Tamarat}.
Under external electric field centrosymmetric molecules such as
DBATT usually gain permanent dipole moments due to distortions
induced by the surrounding solid matrix.
This leads to a linear contribution to the Stark shift, which is
usually much stronger than the quadratic contribution.
In disordered matrices such as in polymers the permanent electric
dipole moment can be as large as 1D, and is around 0.3D in
n-Hexadecane Shpol'skii matrix (corresponding to $\sim 3$ MHz/(kV/m)) \cite{Stark_shift}.
This allows one to use single molecules as highly sensitive probes
of their nanoenvironment and of local electric fields \cite{Faure}.

By setting a bias voltage between the nanotube (NT)
and the conducting (and transparent) substrate, it is possible to
generate an electric field between the NT tip and the substrate
at the limit of the discharge field $E_c=10^7$ V/m \cite{APL2005},
with a very large radial gradient of the order of $E_c/R=10^{16}$ V/m$^2$,
where $R\approx 1$ nm is the NT radius \cite{supplement}.
Any small displacement of the NT will thus induce a large modulation of the
molecular Stark shift, allowing an efficient mechanical/optical transduction.
For typical molecules used in SMS experiments this gives an expected
electric coupling constant $\alpha \approx 10^{19}$ Hz/m,
four orders of magnitude larger than the magnetic coupling constant
observed in Ref. \cite{Arcizet}.

In order to show the capabilities of this detection technique
we provide explicit predictions for the luminescence spectrum and the
second order photon correlation function [$g^{(2)}(t)$] of the molecule
coupled to the oscillator.
Specifically we show:
(i) The modification of the molecular fluorescent excitation spectrum.
(ii)  The possibility of real time displacement detection.
(iii) The relation between $g^{(2)}(t)$ and the oscillator spectrum,
allowing to use the detection scheme in the non-adiabatic regime.
(iv) The possibility of the back action cooling of the oscillator.

\paragraph{Model}
Our description of the proposed experiment is based on the Hamiltonian
$ H=H_M+H_L+H_{phot}+H_{osc}+H_\xi $.
Here $H_M=\hbar\omega(x)\hat{n}$ describes the electronic state of the molecule,
modeled as a TLS,  $\{|1\rangle,|2\rangle\}$, with
$\hat{\pi}=|1\rangle\langle 2|$, and $\hat{n}=|2\rangle\langle 2|$.
The TLS energy splitting is
$
\omega(x)
=
\omega_0+\Delta\vec{\mu}\cdot \vec{E} (x)/\hbar
\approx
\omega_0+\Delta\vec{\mu}\cdot \vec{E} (0)/\hbar+\alpha x$,
where $\omega_0$ is the bare level splitting of the molecule,
$\Delta\vec{\mu}=\langle 2|\vec{\mu}|2\rangle-\langle 1|\vec{\mu}|1\rangle$
is its permanent electric dipole moment,
$\alpha=\partial_x[\Delta\vec{\mu}\cdot \vec{E} (x)]|_{x=0}/\hbar$,
and $\vec{E}(x)$ is the electric field at the molecule that
depends on the displacement $x$ of the nanotube tip from its
equilibrium position.
For simplicity we will focus on a single oscillator mode, thus $x$ is a scalar.
The term $H_L=\Omega_L\hat{\pi} e^{i\om_L t}+h.c$ describes
the coupling of the TLS to a laser field of frequency $\omega_L$
and with amplitude parametrized by the Rabi frequency $\Omega_L$.
The coupling with the photon vacuum environment is given by
$H_{phot}$ and leads to a decay rate $\Gamma$ of the TLS.
Finally $H_{osc}=p_x^2/(2m)+m\om_M^2 { x}^2/2$ and  $H_\xi$
describe the selected mode of the mechanical oscillator (with resonating
frequency $\omega_M$ and effective mass $m$) and
its coupling to the a thermal environment of temperature $T$
inducing a damping rate $\gamma$ and stochastic force fluctuations.

Under the usual assumptions of a short memory photon vacuum \cite{Loudon}
the equations of motion for the
reduced density matrix elements $\s_{12}=\s_{21}^*=\langle\hat{\pi}\rangle$
and $\s_{22}=1-\s_{11}=\langle \hat{n}\rangle$ of the TLS
take the form of Bloch equations:
\bsub\bn \frac{d\s_{12}(t)}{dt}&=&-i\left[\delta+\alpha x(t)\right]\s_{12}(t)-\frac{\Gamma}{2}\s_{12}(t)+\nonumber \\
&&i\Omega_L\left[2\s_{22}(t)-1\right],\\
\frac{d\s_{22}(t)}{dt}&=&-2\Omega_L\Im\left[\s_{12}(t)\right]-\Gamma\s_{22}(t),
\en\label{eq:TLS_osc}\esub
with $\delta=\om(0)-\om_L$ the detuning.

On the same grounds, the dynamics of the average of $x$ after tracing out
the environment degrees of freedom is described by the Langevin equation:
\bn
\frac{d^2x(t)}{dt^2}+\gamma \frac{dx(t)}{dt}+\om_M^2 x(t)=\frac{\xi(t)}{m}+f_{ba}(t) .\label{eq:osc_back}
\en
Here the force $\xi(t)$ is a gaussian fluctuating field with
$\langle \xi(t)\rangle=0$ and $\langle \xi(t) \xi(t') \rangle=2m\gamma k_B T\delta_D(t-t')$
($k_B$ is the Boltzmann constant and $\delta_D(t)$ is the Dirac delta function).
We restrict to a classical description of the oscillator ($k_B T \gg \hbar \omega_M$)
that is the more relevant case for the typical experimental conditions.
The last term in Eq. (\ref{eq:osc_back}) describes the back-action force acting on the oscillator:
$m f_{ba}(t) =\hbar \alpha \sigma_{22}(t)$.

We begin by assuming that the back-action force is negligible.
The validity of this assumption will be discussed later.

\paragraph{Luminescence Spectrum}

Resonant fluorescent excitation spectrum is proportional to the population of the TLS excited state $\s_{22}$.
Solving Eqs. (\ref{eq:TLS_osc}) to order $\Omega_L^2$ for $\Omega_L/\Gamma \ll 1$
we obtain
\bq \s_{22}(t)= \Omega_L^2
\left|\int_{-\infty}^t dt_1 e^{-i\left(\delta-\frac{i}{2}\Gamma\right)(t-t_1)-i\alpha\int_{t_1}^td\tau x(\tau)}\right|^2.
\label{eq:nonadiab}\eq
Eq.  (\ref{eq:nonadiab}) has to be averaged ($\langle ...\rangle_x$) over the displacement fluctuations describled by
Eq. (\ref{eq:osc_back}) to obtain the stationary population \cite{supplement}.
For small oscillator damping, $\gamma \ll \om_M, \Gamma$, this gives
\bn
\langle \s_{22}\rangle_x =
\Omega_L^2\frac{\Gamma+\Gamma_\phi}{\Gamma}  \sum_{n=-\infty}^{+\infty}\frac{ e^{-\zeta^2} I_n\left(\zeta^2\right)} {\left(\delta+n\om_M\right)^2+\frac{1}{4}\left(\Gamma+\Gamma_\phi\right)^2}, \label{eq:rf_I}
\en
where $I_k(z)$ are the modified Bessel functions,
$\zeta^2=\alpha^2\langle x^2\rangle_x/\om_M^2$ is an effective
TLS-oscillator dimensionless coupling constant involving the temperature
($\langle x^2\rangle_x=k_B T/m \om_M^2$) and
$\Gamma_\phi=2 \gamma \zeta^2$
is the TLS mechanically-induced dephasing rate.
%

%
%%%%%%%%%%%%%%%%%%%%%%%%%%%%%%%%%%%
%
%						Figure 2
%
%%%%%%%%%%%%%%%%%%%%%%%%%%%%%%%%%%%
%
%
\begin{figure}[tbp]
  \includegraphics[width=3in]{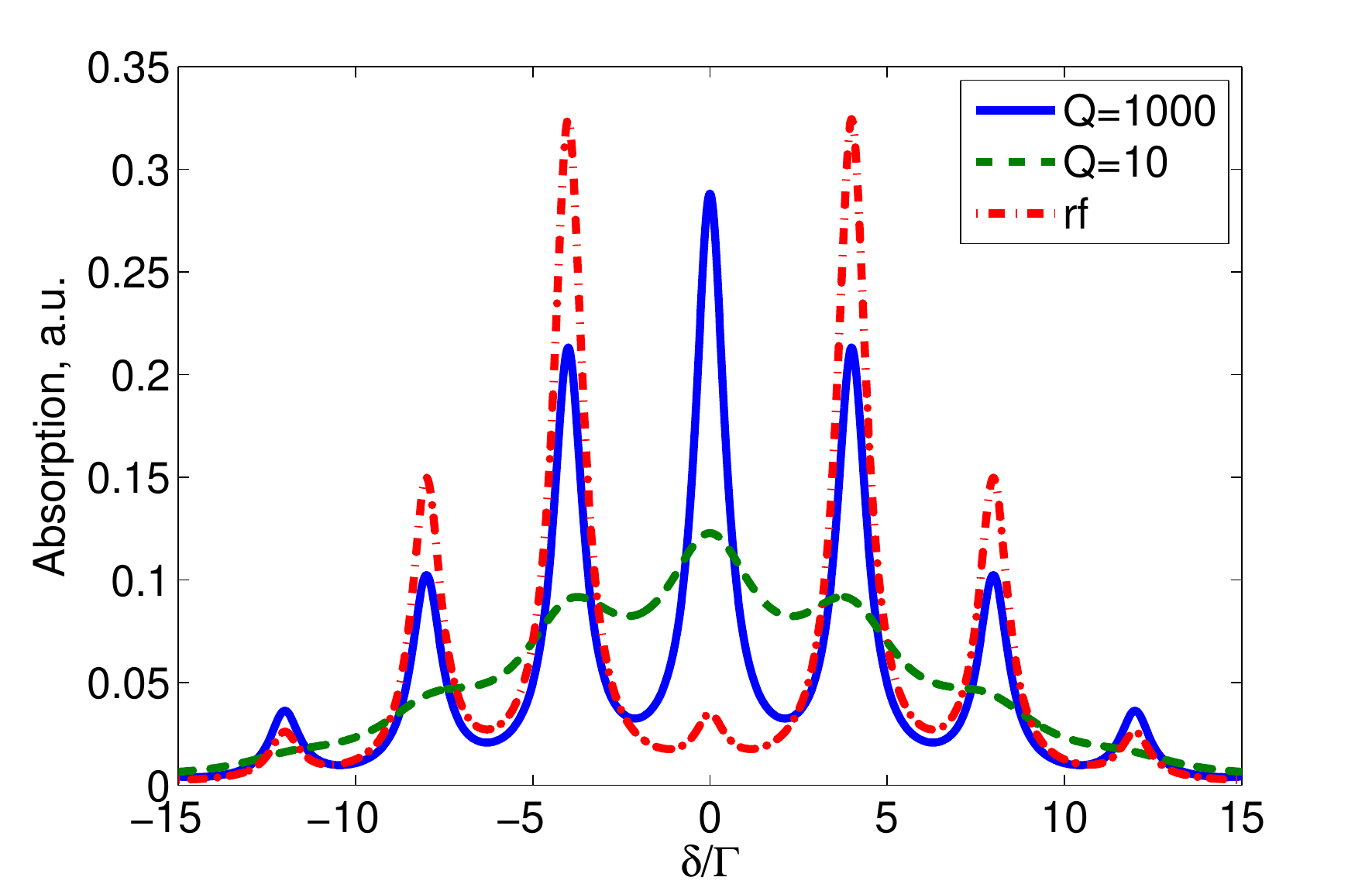}
  \caption{(Color online) Luminescence spectra for the TLS interacting with the
oscillator for $\zeta=1.5$ and $\om_M=4\Gamma$.}\label{fig:osc_G}
\end{figure}
Fig. \ref{fig:osc_G} shows a typical luminescence spectrum obtained from Eq. \ref{eq:rf_I}.
If not otherwise stated, in the Figures and estimates below we choose the
following values of the parameters, $\Gamma/(2\pi)=10$MHz, $\Omega_L=.1\Gamma$,
$m=10^{-21}$kg, and the mechanical oscillator quality factor $Q=\om_M/\gamma=10^3$.
The main feature of the spectrum is the appearance of the sideband peaks at
frequencies $\delta \pm n\om_M$, with $n$ integer.
Their intensity is controlled by $\zeta$ through the Bessel functions.
For $\zeta\ll 1$ a measure of the oscillator fluctuation can be obtained directly
from the ratio of the height of two subsequent peaks
${I_{n+1}\left(\zeta^2\right)}/{I_n\left(\zeta^2\right)}={\zeta^2}/{2(n+1)}$.
Noteworthy that for low quality factors (cf. $Q=10$ shown in Fig. \ref{fig:osc_G})
the peaks significantly broaden due to the mechanically induced dephasing \cite{supplement}.

It is instructive to compare this result to the case of a molecule driven
by a perfectly periodic electric signal, such as a radio frequency (rf) field \cite{Brunel,Tamarat}.
Replacing  $\alpha x(t)$ with $\zeta_{rf} \om_M\cos(\om_M t)$ in Eq.~(\ref{eq:nonadiab}), where $\zeta_{rf}$ is the dimensionless strength of the rf field, one obtains:
\bq
\overline{\s_{22}}
=
\Omega_L^2\sum_{n=-\infty}^{+\infty}
\frac{J^2_n\left(\zeta_{rf}\right)}
{\left(\delta+n\om_M\right)^2+\frac{1}{4} \Gamma^2}.
\label{eq:rf_J}
\eq
The overline indicates averaging over the phase of the rf field.
For $\zeta_{\rm rf}\ll 1$ the spectra due to Eqs. (\ref{eq:rf_J}) and (\ref{eq:rf_I})
essentially coincide for $\zeta_{\rm rf}=\zeta\sqrt{2}$.
However, for $\zeta_{\rm rf}\gg 1$, as a function of $n$
$J^2_n(\zeta_{\rm rf})$ oscillates, which results in the maximum of
intensity shifting to the satellite peaks,
whereas $I_n(\zeta_{\rm rf}^2/2)$ monotonically decreases.
A simple interpretation of this fact can be obtained by
averaging Eq. (\ref{eq:rf_J}) over a gaussian distribution of $\zeta_{\rm rf}^2$ \cite{supplement}.
One finds that the result is very similar to Eq. (\ref{eq:rf_I}), apart from the
broadening of the peaks.
The spectrum (\ref{eq:rf_I}) could thus be viewed with good accuracy as the ensemble average of
many rf sources with random phases and a gaussian distribution of the squared amplitude.

\paragraph{Real Time Position Measurement}

In the adiabatic limit, $\om_M\ll\Gamma$, the sideband peaks merge into a single peak, whose maximum shifts proportionally to the oscillator displacement.
Treating $\alpha x(t)$ as a time-independent shift in the Bloch equations, one obtains
\bn
\s_{22}(t)
=
\frac{\Omega_L^2}{\left[\delta+\alpha x(t)\right]^2+\frac{\Gamma^2}{4}+2\Omega_L^2}.
\label{eq:adiabatic}
\en

It is thus possible to monitor the real time position of the
oscillator by measuring the variation of the luminescence intensity for a given value
of the laser frequency:
$I[x(t)] = \eta \Gamma \s_{22}(t)$,
where $\eta$ is the quantum efficiency of the detector.
The change in the intensity in response to the oscillator displacement is
 \bq G =
 \left.\frac{\partial I[x(t)]}{\partial x(t)}\right|_{x(t)=0}=
-\frac{2 \eta \Gamma \Omega_L^2 \delta \alpha}
{\left(\delta^2+\frac{\Gamma^2}{4}+2\Omega_L^2\right)^2}.
\label{Geq}
\eq

Important characteristic of a displacement detector is its ``sensitivity,"
i.e. the output noise referred back to the input parameter \cite{Zant}:
$\bar{S}_{xx}=\sqrt{S_{II}(0)/G^2}$, where $S_{II}(0)$ is the spectrum
of the intensity fluctuations at zero frequency.
In the adiabatic regime one can treat these fluctuations as a Poissonian
process of emitting single photons, for which
$S_{II}(0)=I[x=0]$.
We thus obtain
\bq \bar{S}_{xx}=\frac{1}{2\alpha \sqrt{\eta\Gamma}}\frac{\left(\delta^2+\frac{\Gamma^2}{4}+\Omega_L^2\right)^{\frac{3}{2}}}
{\Omega_L\left|\delta\right|}.\eq
This expression takes minimum value at $|\delta|=\Omega_L\sqrt{2}=\Gamma/2$.
Taking $\alpha=10^{19}$Hz/m and $\eta=0.01$ we obtain
$\bar{S}_{xx}=\sqrt{27/8}\sqrt{\Gamma/\eta}/\alpha\sim 10 \rm{fm}/\sqrt{\rm{Hz}}$.
This is on par with the best values of sensitivity obtained by
non-interferometric displacement detection methods \cite{sensitivity}.
With respect to the latter, the important advantage of the proposed setup
is its ``open'' nature, which allows a simpler integration of the mechanical
resonator with others systems.

Another essential characteristic of a detector is its
resolution, $\Delta x=\bar{S}_{xx}/\sqrt{\Delta t}$, {\em i.e.}
the smallest displacement that can be measured \cite{Zant}.
Here $\Delta t$ is the measurement time, which
in the case of a damped harmonic oscillator, Eq. (\ref{eq:osc_back}),
can be as long as $\gamma^{-1}=Q/\om_M$.
Thus, for the typical nanotube oscillator frequencies in the range of
$1$kHz-$10$MHz and quality factor $Q=10^3$ the resolution will vary from $10$fm to $1$pm,
which again matches the values obtained by state of the art non-interferometric detection devices \cite{Zant}.
The device can be also used to perform real time dispacement detection, without assuming
a slowly damped  oscillatory behavior of $x(t)$.
In order to have a sufficient signal to noise ratio in this case the limitation
on the oscillator frequency is even more severe giving
$\om_M \ll [\Gamma/(\alpha\bar{S}_{xx})]^2=1$MHz,
where we used also the condition of validity of the linear expansion $\alpha x \ll \Gamma$.

\paragraph{Second order photon correlation function}
In the most common situations, where one is interested in statistical characteristics,
rather than the exact time evolution of the oscillator, one can resort to the measurement
of second order photon correlation function [$g^{(2)}(t,t')$] \cite{Loudon}, that is obtained by
averaging over many two-photon events.
We will show that this allows to improve the frequency resolution even beyond $\Gamma$.
The measured signal for $g^{(2)}(t,t')$ reads:
\bq
g^{(2)}(t,t')=\frac{
\left\langle \left\langle
\hat{\pi}^\dagger(t')\hat{\pi}^\dagger(t)\hat{\pi}(t)\hat{\pi}(t')\right\rangle \right\rangle_x
}
{\left\langle \sigma_{22}(t')  \right\rangle_x
\left\langle \sigma_{22}(t) \right\rangle_x}.
\eq
At equal times $g^{(2)}(t,t)$ vanishes (${\hat\pi}^2=0$), since two photons cannot
be emitted by a single molecule simultaneously (anti-bunching).
For large times ($\tau=t-t' \gg\Gamma^{-1}$) the photon emission events are no more quantum
correlated and the degree of second order coherence reduces to the classical correlation function for the excited state occupation number:
$g^{(2)}(t,t')\approx \langle \s_{22}(t)\s_{22}(t')\rangle_x/\langle \s_{22}\rangle_x^2$.
For vanishing coupling to the oscillator $g^{(2)}(t,t')$ and at lowest order in
$\Omega_L$ exhibits decaying oscillations at the detuning frequency $\delta$:
\bq g^{(2)}(\tau)=1+e^{-\Gamma\tau}-2\cos\left(\delta\tau\right)e^{-\frac{\Gamma\tau}{2}}. \label{eq:g2_nonint}
\eq

\begin{figure}[tbp]
%  \includegraphics[width=3in]{g2_resonance}\\
  \includegraphics[width=3in]{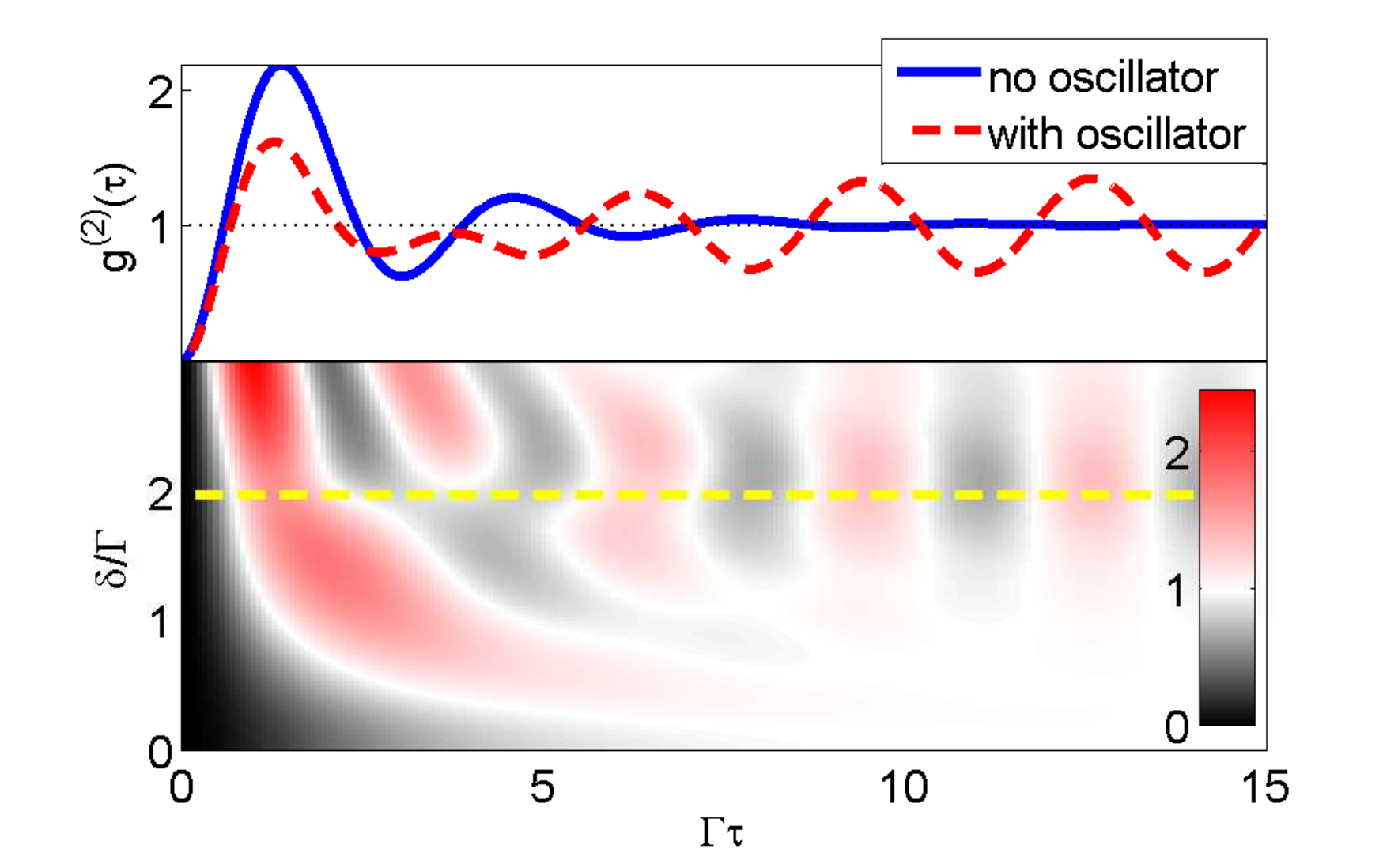}\\
  \caption{(Color online) Top: $g^{(2)}(\tau)$ as a function of time for $\delta=\om_M=2\Gamma$ without/with coupling to the oscillator. Bottom: colorplot of $g^{(2)}(\tau)$ as a function of time and detuning. The dash line markes $\delta=\om_M$. Parameters: $\om_M=2\Gamma$, $\zeta=0.15$, $Q=10^3$.}\label{fig:colorplot}
\end{figure}

It is possible to obtain analytical expressions for $g^{(2)}(t,t')$ in presence of the oscillator
at lowest order in $\Omega_L/\Gamma\ll 1$ and $\zeta\ll 1$ \cite{supplement}.
The result is in general complex, but in some regimes it has simple and interesting interpretation.
For $\delta=\omega_M$ we find that the free oscillations described by
Eq. (\ref{eq:g2_nonint}) are in counter-phase with those induced by the oscillator
leading to a  reduction of the amplitude of the oscillations for $\tau \Gamma \sim 1$
(cfr. Fig. \ref{fig:colorplot}).
For $\Gamma\tau \gg 1$ the Rabi oscillations of Eq. (\ref{eq:g2_nonint}) die out and
the remaining time dependence is all due to the time-correlations induced by the
oscillator $\langle \sigma_{22}(t)\sigma_{22}(t')\rangle_x$.
This is particularly interesting in the case of fast oscillators $\om_M \gg \Gamma \gg \gamma$
for which we find
\bq
g^{(2)}(\tau)
=
1+\frac{4\zeta^2\om_M^2}{\Gamma^2}e^{-\frac{\gamma\tau}{2}}\cos\left(\om_M\tau-Q^{-1}\right).
\eq
The amplitude of $g^{(2)}$ gives thus a direct measure of the average amplitude fluctuation of the oscillator.
In the opposite limit, $\om_M\ll\Gamma$, one obtains
\bq
g^{(2)}(t,t')
=
1+\left(\frac{G}{I}\right)^2 \left\langle x(t)x(t')\right\rangle_x,
\label{eq:nt2nt1}
\eq
valid for an arbitrary oscillator correlation function.

\paragraph{Current-induced dissipation}

The strong coupling between the NT and
the TLS is possible due to the large electric field between the
NT tip and the conducting substrate \cite{supplement}.
Oscillations induce thus dissipative electric currents.
One can single out two different contributions from
the currents flowing in the nanotube and in the substrate.
We find that the dominant effect comes from the surface losses and that for typical parameters the increase in the
oscillator damping rate is $1/(z^3\sigma_{sub})$GHz nm$^{3}$ S/m,
where  $\sigma_{sub}$ is the substrate conductivity and
$z$ the distance from the substrate.
Thus even if it is possible to minimize
this contribution by using
higly conducting substrates,
in common situation it is likely that
the quality factor of the oscillator will be reduced.

\paragraph{Back-action and the effective temperature}

Let's now discuss in more details the role of the back-action force.
Its effect is controlled by two parameters, the intensity of the force
(measured by the ratio between the displacement induced
in the equilibrium position of the oscillator $\hbar \alpha/m \omega_M^2$
and $\sqrt{\langle x^2 \rangle_x }$),
and the frequency and duration of its fluctuation.
In order to estimate its average effect taking both parameters
into account we resort to the linear back-action theory \cite{linearbackaction}.
Within this approach action of the TLS on
the oscillator can be obtained by evaluating the fluctuation
spectrum of the force ($\hbar \alpha \hat n$) in the absence of the mechanical oscillator.
This is proportional to
\bq
S_{nn}(\om)=\int dt e^{i\om t}\left\langle \tilde{n}(t) \tilde{n}(0)\right\rangle=
\frac{\Gamma\langle \hat n\rangle}{\left(\delta-\om\right)^2+\frac{\Gamma^2}{4}},
\eq
where $\langle \hat n\rangle=\Omega_L^2/\left(\delta^2+\Gamma^2/4\right)$,
$\tilde{n}=\hat n-\langle \hat  n\rangle$.
For $Q\gg1$ one can show \cite{linearbackaction} that the oscillator fluctuations
can then be fully characterized by an effective temperature ($T_{osc}$) defined by
\bq \coth\left(\frac{\hbar\om_M}{2k_B T_{\rm osc}}\right)=
\frac{\gamma\coth\left(\frac{\hbar\om_M}{2k_B T}\right)+\gamma_1\coth\left(\frac{\hbar\om_M}{2k_B T_{\rm TLS}}\right)}
{\gamma+\gamma_1},\label{eq:cooling}
\eq
where
$T_{\rm TLS}=\hbar\om_M S_{nn}^+/\left[2k_B S_{nn}^-\right]$,
$\gamma_1=\left(\alpha x_{zpf}\right)^2S_{nn}^-$,
$S_{nn}^\pm=S_{nn}(\om_M)\pm S_{nn}(-\om_M)$, and
$x_{zpf}=\sqrt{\hbar/(2m\om_M)}$.
When $\gamma_1 \gg \gamma$ the oscillator temperature is completely determined by $T_{TLS}$,
which in turn is controlled by the detuning, $\delta$.
For positive values of $\delta$, i.e. for the photon energies below the transition energy,
this results in cooling the oscillator down to the temperature $T_{\rm TLS}$
\cite{Clerk_cooling,cooling}.
At the optimal value $\delta=\sqrt{\omega_M^2+\Gamma^2/4}$ we obtain
$T_{\rm TLS} =\hbar\Gamma/(4 k_B)$ for $\om_M\ll\Gamma$ and
$T_{\rm TLS} \approx \hbar \omega_M/[(2 k_B \log(4 \omega_M/\Gamma)]$
in the resolved side-band limit, $\om_M\gg\Gamma$.
Note that unlike in the case of coupling to a photon cavity \cite{Clerk_cooling,cooling},
where one can infinitely increase the number of cavity photons, the population
of the excited state ($\langle \hat n \rangle $) entering linearly the expression for $\gamma_1$
saturates at high laser strengths and remains smaller than one half.

\begin{figure}[tbp]
  \includegraphics[width=3in]{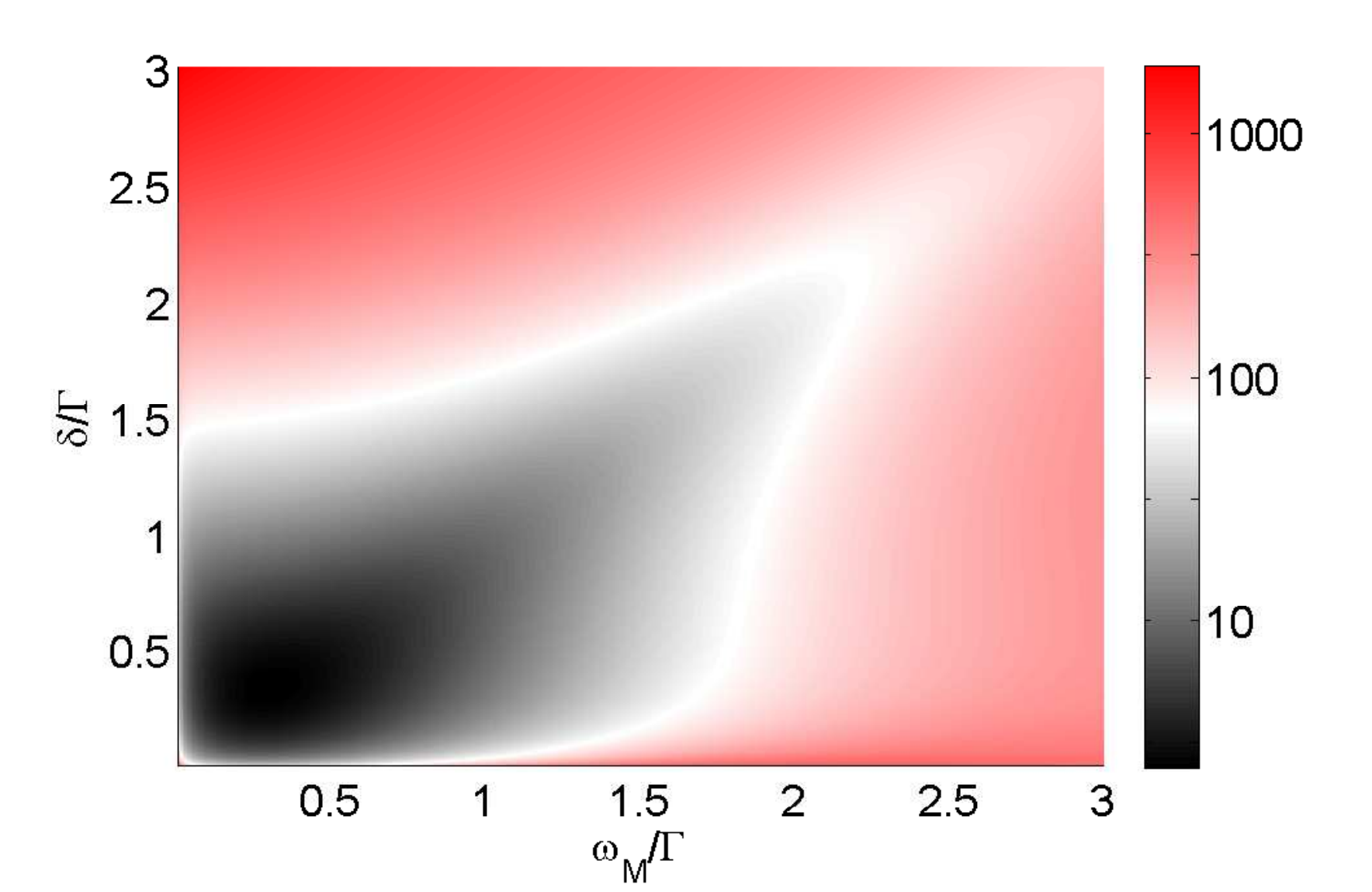}\\
  \caption{(Color online) Ratio $k_B T_{osc}/(\hbar \om_M)$ as the function of the detuning and the oscillator frequency for $\alpha=10^{19}Hz/m$, $T=1K\approx 22$GHz.}\label{fig:Tcolor}
\end{figure}

The effective temperature expression (\ref{eq:cooling})
holds for $\gamma_1 \ll \Gamma$ \cite{Clerk_cooling},
whereas cooling requires that $\gamma_1 \gg \gamma$.
In the resolved side-band limit these conditions result in \cite{supplement}
$\sqrt{\gamma\Gamma/\langle n\rangle} \ll \alpha x_{zpf} \ll\Gamma/\sqrt{\langle n\rangle}$, whereas for $\om_M\ll\Gamma$ the conditions read the same, but with $x_{zpf}$ replaced by $\sqrt{\hbar/(2m\Gamma)}$.
For $\om_M \approx \Gamma$ we thus obtain that the linear response description is correct for $\alpha<10^{19}$Hz/m, {\em i.e.} under all realistic circumstances,
whereas the back-action effects are negligible for $\alpha\ll 10^{17}$Hz/m.
This gives the quantitative condition for discarding $f_{\rm ba}$ in the first
part of this paper.
This also indicates that there is a window of at least two orders
of magnitude of the coupling
constant where back-action can be used to manipulate conveniently the oscillator.
The dependence of the effective temperature on $\delta/\Gamma$ and
$\om_M/\Gamma$ is shown in Fig. \ref{fig:Tcolor}:
The temperature of the oscillator can be reduced by a factor of a 1000
when $\omega_M/\Gamma \approx 0.1-0.4$.

\paragraph{Conclusion}

In this Letter we have shown that single-molecule spectroscopy can be used as an efficient tool for detecting and manipulating nano-oscillators, such as carbon nanotubes attached to an AFM tip and interacting with the molecule via electrostatic forces.
We analyzed different detection schemes which allow obtaining either real time or averaged information on the oscillator displacement. Estimated very large values of the coupling constant allow reaching high detection efficiency.
Furthermore, the back-action of the molecule on the oscillator can be made sufficiently large to enable strong cooling of the oscillator with the proposed set-up.
We considered in this Letter only the linear part of the TLS-oscillator coupling,
but by properly position the nanotube it is possible to work in a regime where
the dominant coupling is quadratic, leading to the wide range of possibilities
including that of measuring the energy of the oscillator \cite{YaleGroup}.
Single-molecule spectroscopy appears thus as a promising new technique
for detecting displacement and exploring strong coupling regimes of
a driven two-level system with a nano-mechanical oscillator.

\paragraph{Acknowledgements}

F. P. and V.P. acknowledge financial support from the {\em R\'egion Aquitaine} through contract
n. 20101101013 and the French ANR through contract QNM n. 040401.
B. L. acknowledges the financial support from the ANR end the ERC.

\end{document}

\title{Single molecule detection of nanomechanical motion: Supplementary material}
\author{Vadim Puller$^1$}
\author{Brahim Lounis$^2$}
\author{Fabio Pistolesi$^1$}
\affiliation{
$^1$
Univ. Bordeaux, LOMA, UMR 5798, F-33400 Talence, France.\\
CNRS, LOMA, UMR 5798, F-33400 Talence, France.\\
$^2$
Univ Bordeaux, LP2N,  F-33400 Talence, France\\
Institut d'Optique and CNRS, UMR 5298, Talence, France
}

\begin{abstract}
{We provide here more detailed derivations of some of the results presented in the main manuscript \cite{Puller}}
\end{abstract}
\date{\today}
%\pacs{73.23.-b, 73.23.Hk, 73.63.Kv}

 \maketitle

\section{Spectrum of the TLS coupled to an oscillator}
Solving Eqs. (1) and (2) of Ref. \onlinecite{Puller} to order $\Omega_L$ for $\Omega_L/\Gamma \ll 1$ (and neglecting the back-action) we obtain
\bsub\bn
\s_{12}(t) &\approx& -i\Omega_L \int_{-\infty}^t dt_1e^{-i\left(\delta- \frac{i}{2}\Gamma\right)\left(t-t_1\right)}
e^{- i\alpha\int_{t_1}^t d\tau x(\tau)},\label{eq:s12}\\
\s_{22}(t) &=& -2\Omega_L\Im\left[\int_{-\infty}^tdt_1 \s_{12}(t_1) e^{-\Gamma(t-t_1)}\right].\label{eq:s22}
\en\label{eq:orderOmega}\esub
Substituting Eq. (\ref{eq:s12}) into Eq. (\ref{eq:s22}) and using algebraic manipulations one can show that $\s_{22}(t)$ is given by Eq. (3) of Ref. \onlinecite{Puller},
which is convenient for the analysis of the adiabatic limit and the degree of second order coherence.
For now we focus on Eqs. (\ref{eq:orderOmega}).
One can use the Gaussian properties of the stochastic noise, $\xi(t)$, (and therefore also of $x(t)$) to carry out the averaging over the oscillator position fluctuations:
\bq  \left\langle e^{- i\alpha\int_{t_1}^td\tau x(\tau)}\right\rangle=e^{\Phi(t-t_1)},\eq
where we took into account that $\langle x(t)\rangle=0$, whereas
\bq \Phi(t)=-\frac{\alpha^2}{2}\int_0^td\tau_2\int_0^td\tau_1\langle x(\tau_2)x(\tau_1)\rangle,\eq
with the oscillator correlation function given by
\bq \langle x(t)x(t')\rangle=\frac{\langle x^2\rangle}{\bar{\om}_M}\Im\left[\lambda e^{\lambda^*|t-t'|}\right]=
\langle x^2\rangle e^{-\frac{\gamma|t-t'|}{2}}\left\{\cos[\bar{\om}_M(t-t')]+ \frac{\gamma}{2\bar{\om}_M}\sin[\bar{\om}_M|t-t'|]\right\}.\label{eq:osc_corr}\eq
Here $\lambda=-\gamma/2+i\bar{\om}_M$, $\bar{\om}_M=\sqrt{\om_M^2-\gamma^2/4}$, and $\langle x^2\rangle=k_B T/(m\om_M^2)$ is the level of the oscillator fluctuations as determined by the equipartition theorem.

We now have
\bq
\langle \s_{22}\rangle =\frac{2\Omega_L^2 e^{-(z+z^*)}}{\Gamma}
\Re\left[\int_0^{+\infty} \!\!\!\! \!\!\!\! dt e^{-i\left[\delta-\frac{i}{2}\left(\Gamma+\Gamma_\phi\right)\right]t +ze^{\lambda t}+z^* e^{\lambda^* t}}\right],\label{eq:sigma22}
\eq
where
$\Gamma_\phi=2\gamma\zeta^2$
is the dephasing rate of the TLS induced by the mechanical noise,
$z=\zeta^2(\lambda^*)^3/(2i\bar{\om}_M\om_M^2)$,
and
$\zeta^2=\alpha^2\langle x^2\rangle/\om_M^2$ is the
dimensionless coupling strength between the oscillator and the TLS.

Expanding the integrand of Eq. (\ref{eq:sigma22}) in series in $ze^{\lambda t}$, $z^* e^{\lambda^* t}$ one can carry out the integration with the result
\bq
\langle \s_{22}\rangle =\frac{2\Omega_L^2 e^{-(z+z^*)}}{\Gamma}
\Re\left\{\sum_{n,m=0}^{+\infty}\frac{z^n\left(z^*\right)^m}{n!m!} \frac{1}{i\left[\delta+(m-n)\om_M\right]+\frac{1}{2}\left[\Gamma+\Gamma_\phi+(n+m)\gamma\right]}
\right\}.\label{eq:doublesum}
\eq
This equation was used to obtain Fig. 2 of Ref. \onlinecite{Puller}.

In the limit of the high oscillator quality factor, $Q=\om_M/\gamma\gg 1$,
we have $z\approx\zeta^2/2$ and $\lambda\approx \om_M$.
Then $ze^{\lambda t}+z^* e^{\lambda^* t}\approx \zeta^2\cos(\om_M t)$, and can use the expansion in terms of the modified Bessel functions \cite{AS}
\bq e^{z\cos\theta}=\sum_{k=-\infty}^{\infty}I_k(z)e^{ik\theta}\eq
to obtain Eq. (4) of Ref. \onlinecite{Puller}.
Comparing with Eq. (\ref{eq:doublesum}) we see that this corresponds to assuming that $\gamma\ll\om_M$ and $(n+m)\gamma\ll\Gamma+\Gamma_\phi$, i.e. the approximation is valid when one can neglect the satellites with indices higher than $(\Gamma+\Gamma_\phi)/\gamma$.

\section{Correspondence between oscillator and ac spectra}
Let us consider a molecule driven by an undamped oscillator with a constant amplitude $A$:
$x(t)=A\cos\left[\om_M t+\phi\right]$.
The resulting spectrum is given by Eq. (6) of Ref.\onlinecite{Puller},
where the argument of the Bessel function is
$\zeta_{rf}=\alpha A/\om_M$.
We now will average this spectrum over the thermal distribution of the oscillator energies
\bq W_E\left(E\right)=\frac{1}{k_B T}e^{-\frac{E}{k_B T}}.\eq
The oscillator energy is related to its amplitude as $E(A)=m\om_M^2A^2/2$, which allows us to write the probability distribution for the amplitude
\bq W_A\left(A\right)=
\frac{d E(A)}{dA}W_E\left[E\left(A\right)\right]=
\frac{A}{\langle x^2\rangle_x}e^{-\frac{A^2}{2\langle x^2\rangle_x}}.\eq
(Note that $\langle A^2\rangle_x=2\langle x^2\rangle_x$.)

We now evaluate the average
\bq \left\langle \left[J_n\left(\frac{\alpha A}{\om_M}\right)\right]^2\right\rangle=
\int_0^{+\infty}dA W_A\left(A\right) \left[J_n\left(\frac{\alpha A}{\om_M}\right)\right]^2=
e^{-\langle \zeta_{rf}^2\rangle_x} I_n\left(\langle\zeta_{rf}^2\rangle_x\right),\eq
where $\langle\zeta_{rf}^2\rangle_x=\alpha^2\langle x^2\rangle_x/\om_M^2$, and we have used the relation\cite{Gradshtein}
\bq \int_0^{+\infty}dx e^{-\rho^2 x^2} x J_p(\alpha x)J_p(\beta x)=
\frac{1}{2\rho^2}e^{-\frac{\alpha^2+\beta^2}{4\rho^2}} I_p\left(\frac{\alpha\beta}{2\rho^2}\right).\eq
The resulting averaged spectrum coincides with the one given by Eq. (5) of Ref. \onlinecite{Puller} (at zero damping $\Gamma_\phi=0$).

\section{Degree of second order coherence}
\subsection{Calculating the degree of second order coherence, $g^{(2)}(\tau)$}
The usual way to obtain $g^{(2)}(\tau)$ [Eq. (10) of Ref. \onlinecite{Puller})] is by using the quantum regression theorem,\cite{Loudon,Cohen-Tannouji} which relates $g^{(2)}(\tau)$ to the evolution of the average dipole moment and the excited state occupation number.
This is done by deriving equations for the correlation functions
\bq C_{12}(t,t')=\langle \hat{\pi}^\dagger(t')\hat{\pi}(t)\hat{\pi}(t')\rangle,\textrm{ }
C_{22}(t,t')=\langle \hat{\pi}^\dagger (t')\hat{\pi}^\dagger (t)\hat{\pi}(t)\hat{\pi}(t')\rangle,\eq
using the same approximations as those that were used in deriving the Bloch equations [Eqs. (1) in Ref. \onlinecite{Puller}]:
\bsub\bn \frac{\partial}{\partial t} C_{12}(t,t')&=& -i\left[\delta +\alpha x(t)\right] C_{12}(t,t')-\frac{\Gamma}{2}C_{12}(t,t')+
i\Omega_L\left[2C_{22}(t,t')-\sigma_{22}(t')\right],\\
\frac{\partial}{\partial t} C_{22}(t,t')&=&-2\Omega_L\Im\left[C_{12}(t,t')\right]-\Gamma C_{22}(t,t').\en\label{eq:C12C22}\esub

Eqs. (\ref{eq:C12C22}) are solved in the interval $[t',t]$ with the initial conditions $C_{12}(t',t')=0$, and $C_{22}(t',t')=0$ which readily follow from the definition of these functions.

So far we have not performed the statistical averaging over the thermal fluctuations of the oscillator, $x(t)$. Denoting such averaging by $\langle ...\rangle_x$ we define the degree of second order coherence as
\bq g^{(2)}(t,t')=\frac{\langle C_{22}(t,t')\rangle_x}{\langle \sigma_{22}(t)\rangle_x\langle \sigma_{22}(t')\rangle_x}.\label{eq:g2x}\eq
Implied in this definition is that the Brown-Twiss experiment does not directly measure $g^{(2)}(t,t')$, but rather measures separately  $C_{22}(t,t')$ and $\sigma_{22}(t)$, which are averaged over the oscillator fluctuations after many repeated measurements, and later operated numerically to form $g^{(2)}(t,t')$. Note that after the averaging the degree of second order coherence will depend only on the difference of the two times: $\tau=t-t'$.

\subsection{Degree of second order coherence to order $\zeta^2$}
When solving Eqs. (\ref{eq:C12C22}) to order $\Omega_L^2$, $C_{22}(t,t')$ is given by
\bq
C_{22}(t,t')=
\Omega_L^2 \sigma_{22}(t') \left|\int_{t'}^tdt_1 e^{-i\left(\delta-\frac{i\Gamma}{2}\right)(t-t_1)-i\int_{t_1}^td\tau \alpha x(\tau)}\right|^2,
\label{eq:g2}\eq
where $\s_{22}(t)$ is given by Eq. (3) of Ref. \onlinecite{Puller}.
Thus, $C_{22}(t,t')$ is actually of order $\Omega_L^4$, but the lowest order contribution to the degree of second order coherence, defined by Eq. (\ref{eq:g2x}), will be of zero-th order in $\Omega_L$.

For a Gaussian statistics of $x(t)$ one can explicitly carry out the statistical averaging of $\s_{22}(t)$ and $C_{22}(t,t')$.
In the following we limit ourselves to the limit $\zeta\ll 1$ to
order $\zeta^2$, where most of the integrals can be evaluated analytically.
The expressions given below were used to obtain Fig. 3 of Ref. \onlinecite{Puller}.

After straightforward but tedious calculations one arrives at
\bsub\bn \frac{\left\langle \s_{22}(t)\s_{22}(t')\right\rangle_x}
{\left\langle \s_{22}\right\rangle_x^2}&=&
1+\frac{\delta}{\Gamma}2\Re\left[\frac{I_++I_-}{\delta-\frac{i\Gamma}{2}}\right],\label{eq:densitycorr}\\
\frac{\left\langle \left(\s_{22}\right)^2\right\rangle_x}
{\left\langle \s_{22}\right\rangle_x^2}&=&
1+\frac{\delta}{\Gamma}2\Re\left[\frac{2I_0}{\delta-\frac{i\Gamma}{2}}\right],\\
\left\langle \s_{22}\right\rangle_x &=& \frac{g^2}{\delta^2+\frac{\Gamma^2}{4}}\left\{1+2\Re\left[\frac{\delta+\frac{i\Gamma}{2}} {\delta-\frac{i\Gamma}{2}}\frac{I_0}{\Gamma}\right]\right\},\en\esub
\bn
g^{(2)}(t,t')&=&
\frac{\left\langle \s_{22}(t)\s_{22}(t')\right\rangle_x+
\left\langle \s_{22}^2\right\rangle_x e^{-\Gamma(t-t')}}
{\left\langle \s_{22}\right\rangle_x^2}-\nonumber\\
&&2e^{-\frac{\Gamma(t-t')}{2}+\Phi(t-t')}\left\{\cos\left[\delta(t-t')\right]
\left[\frac{\left\langle \s_{22}(t)\s_{22}(t')\right\rangle_x}
{\left\langle \s_{22}\right\rangle_x^2}+
\Re\left[\frac{i\left(I_++I_--2I_0\right)}
{2\left(\delta-\frac{i\Gamma}{2}\right)}\right]\right]\right.+\nonumber\\
&&\left.\sin\left[\delta(t-t')\right]\Im\left[\frac{4iI_\varphi}
{\delta-\frac{i\Gamma}{2}}+\frac{i\left(I_++I_--2I_0\right)}
{2\left(\delta-\frac{i\Gamma}{2}\right)}+
\frac{2\left(\delta+\frac{i\Gamma}{2}\right)} {\Gamma\left(\delta-\frac{i\Gamma}{2}\right)}
\left(I_+-I_-\right)\right]\right\},\label{eq:g2I}
\en
where we have defined
\bsub\bn
I_0&=&\alpha^2\int_0^{+\infty}d\tau \left\langle x(\tau)x(0)\right\rangle_x
e^{-i\left(\delta-\frac{i\Gamma}{2}\right)\tau},\\
I_\pm &=&\alpha^2\int_0^{+\infty}d\tau \left\langle x(t-t'\pm\tau)x(0)\right\rangle_x
e^{-i\left(\delta-\frac{i\Gamma}{2}\right)\tau},\\
I_\varphi&=&\alpha^2\int_0^{t-t'}d\tau \left\langle x(\tau)x(0)\right\rangle_x,\\
\Phi(t-t')&=&-\frac{\alpha^2}{2}\int_{t'}^td\tau_2\int_{t'}^t d\tau_1
\left\langle x(\tau_2)x(\tau_1)\right\rangle_x .
\en\label{eq:Integrals}\esub
Note that for $t-t'=0$ we have $I_\pm=I_0$,  and $\Phi(0)=0$, so that the coherence function takes zero value, which corresponds to antibunching.

\subsection{Vanishing coupling to the oscillator}
\begin{figure}[tbp]
  \includegraphics[width=3in]{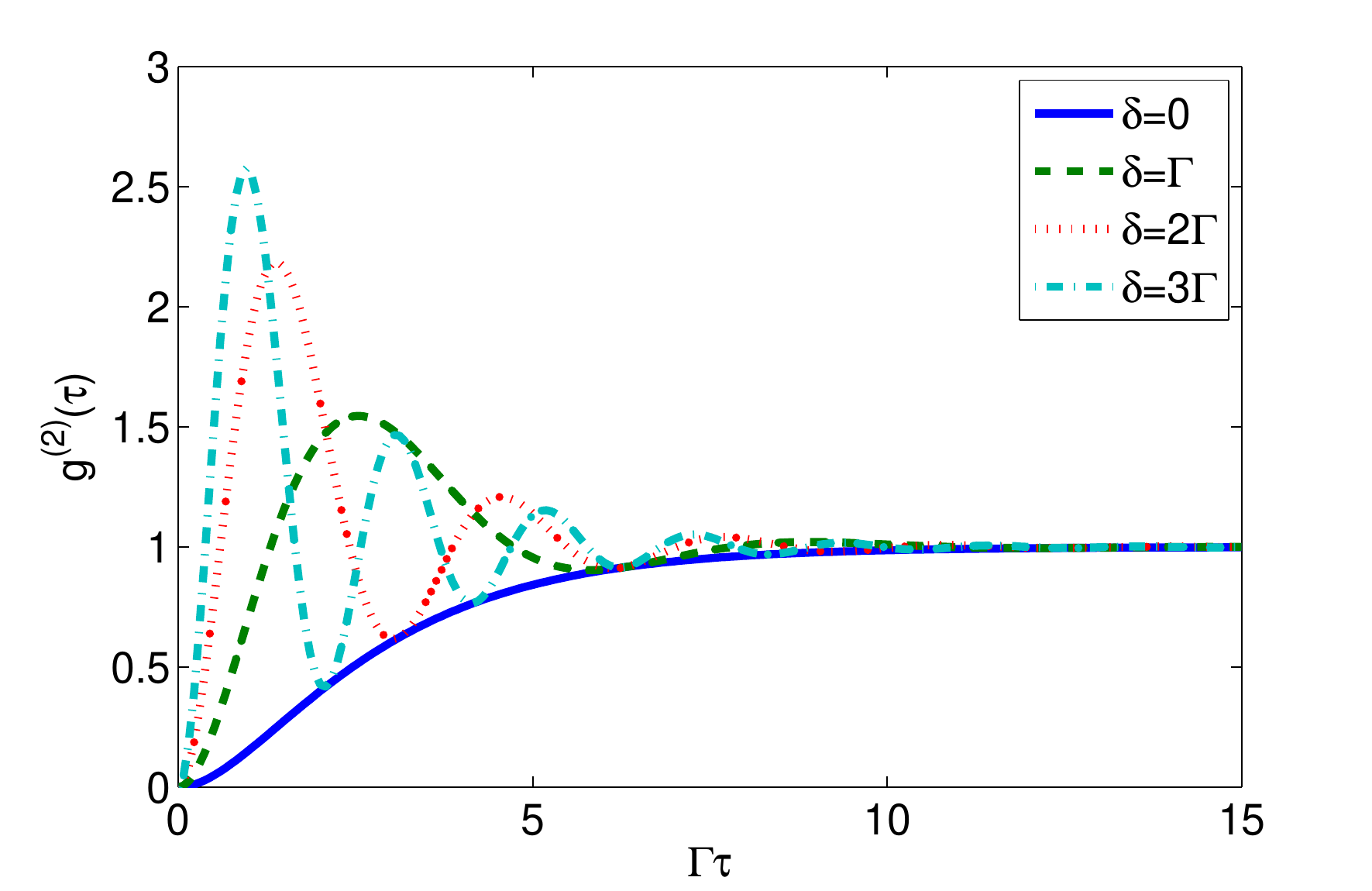}
  \caption{Degree of second order coherence, $g^{(2)}(\tau)$, for a TLS not coupled to an oscillator as a function of time for various values of detuning.}\label{fig:lines}
\end{figure}

\begin{figure}[tbp]
  \includegraphics[width=3in]{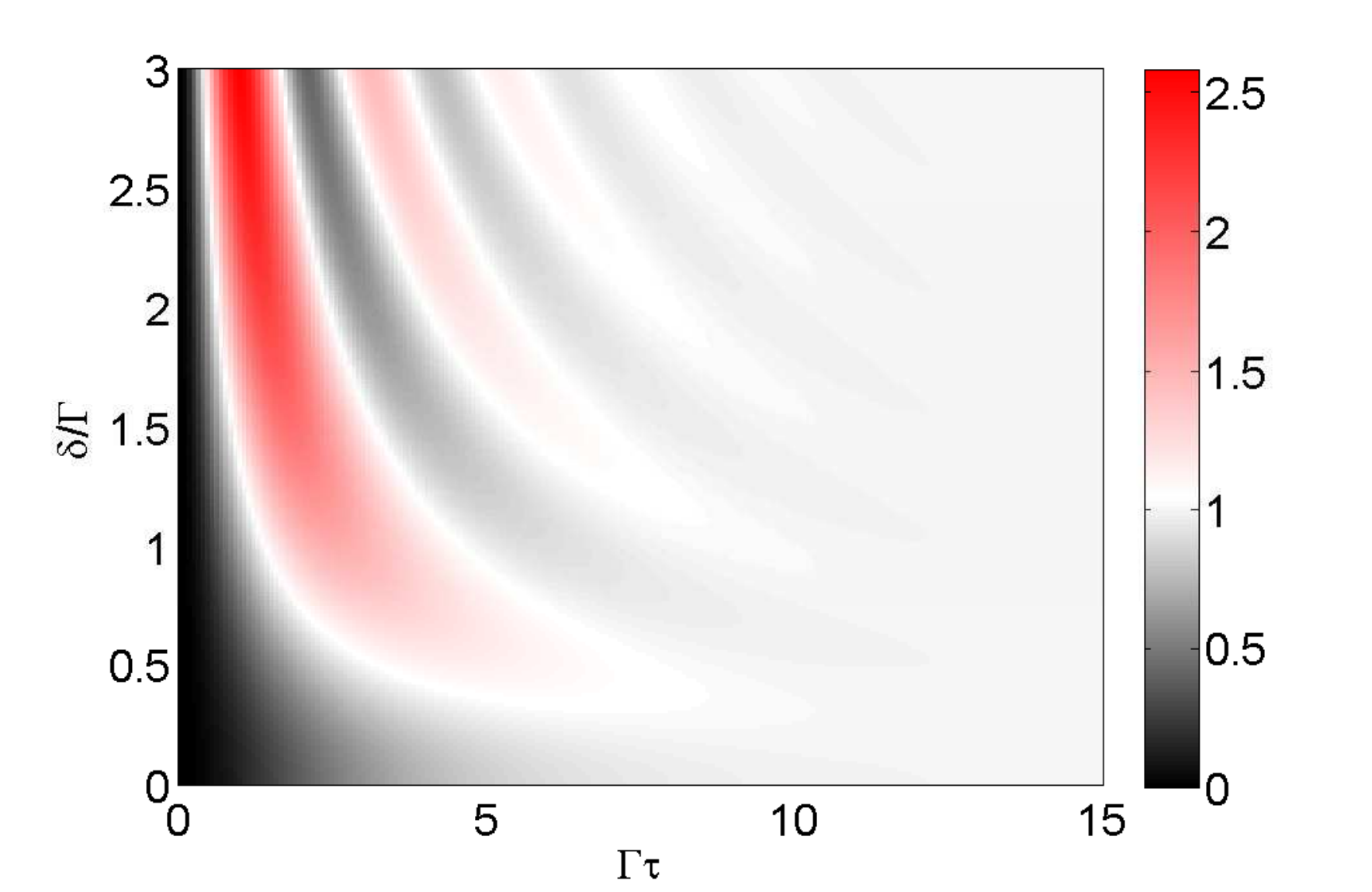}
  \includegraphics[width=3in]{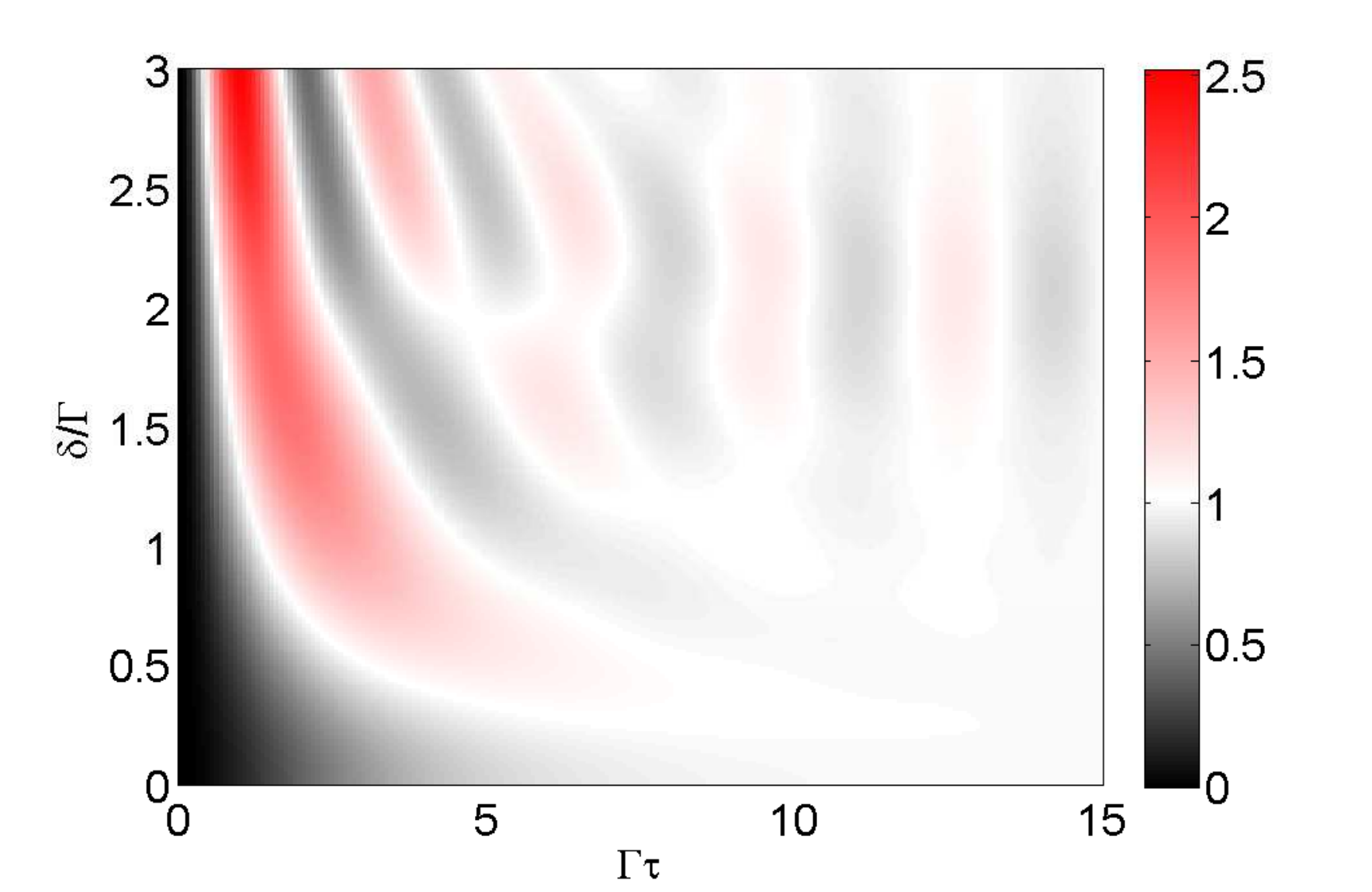}
  \caption{Degree of second order coherence, $g^{(2)}(\tau)$, as a function of time and the detuning for a molecule not coupled to an oscillator (left) and a molecule coupled to an oscillator with $\zeta=.1$, $\om_M=2\Gamma$, and $Q=10^3$ (right).}\label{fig:weak}
\end{figure}

When the coupling to the oscillator vanishes the degree of second order coherence takes the simple form
\bq g^{(2)}(\tau)=1+e^{-\Gamma\tau}-2\cos\left(\delta\tau\right)e^{-\frac{\Gamma\tau}{2}},\eq
which illustrates all the main features of $g^{(2)}(\tau)$ originating from
single molecule luminescence:
i) antibunching, i.e. $g^{(2)}(0)=0$;
ii) unitary limit at long times, $g^{(2)}(\tau\rightarrow +\infty)\rightarrow 1$;
and iii) oscillations with the frequency of detuning at times smaller than $\Gamma^{-1}$. These are illustrated in Fig. \ref{fig:lines} and Fig. \ref{fig:weak} (left panel).

\subsection{Adiabatic limit}
Let us now consider the adiabatic limit, i.e. when the correlation function
$\langle x(t)x(t')\rangle_x$ changes on a time scale much slower than
$\Gamma^{-1}$.
In this case the correlation function can be taken out of the integrals of
Eqs. (\ref{eq:Integrals}) and, for times larger than the lifetime of
the excited state of the molecule, $\tau\gg\Gamma^{-1}$,
we obtain Eq. (11) of Ref. \onlinecite{Puller}.

In the adiabatic limit it is possible to obtain more general results,
not restricted to small values of the effective coupling $\zeta$.
Firstly, for the time intervals $t-t'>\Gamma^{-1}$ relevant to the
adiabatic limit, the degree of second order coherence is reduced to
the correlation function for the excited state population of the molecule:
\bq   g^{(2)}(t,t')\approx\frac{\langle \s_{22}(t)\s_{22}(t')\rangle_x}
{\langle \s_{22}(t)\rangle_x \langle \s_{22}(t')\rangle_x},\eq
where $\s_{22}$ is given by Eq. (7) of Ref. \onlinecite{Puller}.
One can carry out the statistical averaging and express the result in terms of a double integral:
\bsub\bn  \langle \s_{22}(t)\s_{22}(t')\rangle_x &=&
\left(\frac{\Omega_L^2}{\Gamma}\right)^2
\int_{-\infty}^{+\infty}d\tau_2 \int_{-\infty}^{+\infty}d\tau_1 \cos\left[\delta(\tau_2+\tau_1)\right]
e^{-\frac{\Gamma|\tau_2|}{2}-\frac{\Gamma|\tau_1|}{2}-
\frac{\alpha^2\langle x^2\rangle\left(\tau_2^2+\tau_1^2\right)}{2}-
\alpha^2\langle x(t)x(t')\rangle\tau_2\tau_1},\\
\langle \s_{22}(t)\rangle_x &=&
\frac{\Omega_L^2}{\Gamma}
\int_{-\infty}^{+\infty}d\tau\cos(\delta\tau)e^{-\frac{\Gamma|\tau|}{2}-
\frac{\alpha^2\langle x^2\rangle\tau^2}{2}}. \en\label{eq:g2_adiabatic}\esub
Eqs. (\ref{eq:g2_adiabatic}) show that in the adiabatic limit $g^{(2)}(\tau)$
is a non-linear transformation of the oscillator correlation function,
$\langle x(t)x(t')\rangle$.
Thus, although $g^{(2)}(\tau)$ may not always correctly reproduce
the oscillator correlation function, it allows to deduce characteristic
frequencies of the oscillator.

In the limit where the TLS-oscillator coupling is stronger than the molecular linewidth, $\alpha^2\langle x^2\rangle\gg\Gamma^2$, Eqs. (\ref{eq:g2_adiabatic}) can be simplified by neglecting the terms $\Gamma|\tau|/2$ in the exponents.
In this case the intergrals can be evaluated exactly and we arrive at
\bq g^{(2)}(t,t')=
\frac{1}{\sqrt{1-\left(\frac{\langle x(t)x(t')\rangle}{\langle x^2\rangle}\right)^2}} \exp\left[\frac{\delta^2}{\alpha^2\langle x^2\rangle}
\frac{\langle x(t)x(t')\rangle}{\langle x^2\rangle+\langle x(t)x(t')\rangle}\right],\eq
whereas
\bq \left\langle \s_{22}(t)\right\rangle_x=
\sqrt{\frac{2\pi}{\alpha^2\langle x^2\rangle}}\frac{\Omega_L^2}{\Gamma}
e^{-\frac{\delta^2}{2\alpha^2\langle x^2\rangle}},\eq
which is the population of the excited state of the molecule with inhomogeneously broadened line.

\subsection{Fast oscillator}
At times $t-t'\gg\Gamma^{-1}$ the degree of second order coherence,
Eq. (\ref{eq:g2I}), is reduced to the correlation function for the
population of the excited state of the TLS:
\bq \frac{\langle \s_{22}(t)\s_{22}(t')\rangle_x}{\langle \s_{22}\rangle_x^2}=
1+\int_{-\infty}^{+\infty}\frac{d\om}{2\pi}\frac{\alpha^2
S_{xx}(\om)}{\om}\frac{2\delta}{\left(\delta-\om\right)^2+
\frac{\Gamma^2}{4}}\cos\left[\om(t-t')\right],\eq
where we have re-expressed the integrals of Eqs. (\ref{eq:Integrals})
in terms of the oscillator spectrum, defined as
\bq S_{xx}(\om)=
\int_{-\infty}^{+\infty}dt e^{i\om t}\left\langle x(t)x(0)\right\rangle_x.\eq
In case of a linear oscillator with correlation function given by Eq. (\ref{eq:osc_corr})
this spectrum is given by
\bq S_{xx}(\om)=\frac{2\gamma \om_M^2\langle x^2\rangle}{\left(\om^2-\om_M^2\right)^2+\om^2\gamma^2},\eq
and we obtain
\bq \frac{\langle \s_{22}(t)\s_{22}(t')\rangle_x}{\langle \s_{22}\rangle_x^2}=
1+\frac{\zeta^2\delta}{\bar{\om}_M}
\Re\left\{e^{i\bar{\om}_M\tau-\frac{\gamma\tau}{2}}
\left(\bar{\om}_M-\frac{i\gamma}{2}\right)^2
\left[\frac{1}{\left(\bar{\om}_M+\frac{i\gamma}{2}-\delta\right)^2+\frac{\Gamma^2}{4}}-
\frac{1}{\left(\bar{\om}_M+\frac{i\gamma}{2}+\delta\right)^2+
\frac{\Gamma^2}{4}}\right]\right\}.\label{eq:remnantosc}
\eq
In the limit $\delta=\om_M\gg\Gamma$ the second term in this equation can be omitted and we arrive at Eq. (13) of Ref. \onlinecite{Puller}.

For negligibly small oscillator damping Eq. (\ref{eq:remnantosc}) becomes
\bq \frac{\left\langle \s_{22}(t)\s_{22}(t')\right\rangle_x}{\langle \s_{22}\rangle_x^2}=
1+\zeta^2\delta\om_M \left[\frac{1}{(\delta-\om_M)^2+\frac{\Gamma^2}{4}}-\frac{1}{(\delta+\om_M)^2+\frac{\Gamma^2}{4}}\right]
\cos\left[\om_M(t-t')\right]=
1+A\cos\left[\om_M(t-t')\right].\label{eq:remnantosc1}\eq
If we now consider the degree of second order coherence at times
where $e^{-\Gamma\tau/2}\sim \zeta^2$, we can omit the terms proportional to
$\zeta^2 e^{-\Gamma\tau/2}$ and $e^{-\Gamma\tau}$,
but keep the oscillatory term $2e^{-\frac{\Gamma(t-t')}{2}}\cos\left[\delta(t-t')\right]$:
\bq g^{(2)}(t,t')=\frac{\langle \s_{22}(t)\s_{22}(t')\rangle_x}{\langle \s_{22}\rangle_x^2}-2e^{-\frac{\Gamma(t-t')}{2}}\cos\left[\delta(t-t')\right].\label{eq:g2_reduced}\eq
When $\delta=\om_M$ the two terms oscillate at the same frequency but have different signs, i.e. the overall oscillation is suppressed:
\bq g^{(2)}(\tau)=1+\left(A-2e^{-\frac{\Gamma\tau}{2}}\right)\cos\left(\delta\tau\right).\eq
The amplitude of the oscillations thus vanishes at
\bq \tau_0=
\frac{2}{\Gamma}\log\left(\frac{2}{A}\right)=
\frac{2}{\Gamma}\log\left(\frac{\Gamma^2}{2\alpha^2\langle x^2\rangle}\right).\eq
Since zeros of the cosine function ($\om_M\tau=\pi (n+1/2)$) occur at equally spaced intervals, $\tau_0$ is easily identified as the additional crossing of $g^{(2)}=1$.
The exception is the case when $\tau_0$ coincides with one of the zeros of the cosine.
Then the two zeros merge and the oscillations only touch $g^{(2)}=1$.
($g^{(2)}=1$ can be identified as the asymptote of the oscillations at times $\tau\gg\Gamma^{-1}$.)

\section{Estimate of the electric field gradient}
In Ref. \onlinecite{Puller} we estimated that the electric field gradient due to the nanotube by taking the ratio of the maximum electric field (discharge field, $E_c=10^7$V/m) to the nanotube radius ($R\approx1$nm): $E_c/R=10^{16}$V/m${^2}$.
Here we provide more rigorous electrostatic calculations which support this simple estimate.

We treat the nanotube as a perfectly conducting cylinder of radius $R$ with infinitely thin walls, suspended above a conducting surface (substrate).
The boundary of the substrate is specified by $z=0$.
The nanotube axis is taken to coincide with the z-axis, its length is $L$, and the distance between the bottom end of the nanotube and the surface is $z_0$.
The nanotube has potential $V$ in respect to the surface, while the surface itself is assumed to have zero potential.

The electric potential due to the charge in the nanotube and the image charges induced in the substrate is \cite{Rotkin}
\bq \varphi(r,z)=\int_{z_0}^{z_0+L} dz'\lambda(z')\left[F(r,z-z')-F(r,z+z')\right],\label{eq:potential}\eq
where $\lambda(z)$ is the linear charge density along the nanotube (we use cgs units), and the integration kernel is
\bq F(r,z)=\int_0^{2\pi}\frac{d\phi}{2\pi} \frac{1}{\sqrt{r^2+R^2-2rR\cos\phi+z^2}}.\eq
Since we assume that the nanotube is perfectly conducting, all its points have the same potential, i.e.
\bq\varphi(R,z)=V \textrm{ for } z_0\leq z\leq z_0+L,\label{eq:inteq}\eq
which is the integral equation for determining the distribution of the electric charge on the nanotube.
Note that Eqs. (\ref{eq:potential}) and (\ref{eq:inteq}) scale linearly with the potential difference, i.e. increasing/reducing $V$ will proportionally change the electric field affecting the molecule.

\begin{figure}[tbp]
  \includegraphics[width=3in]{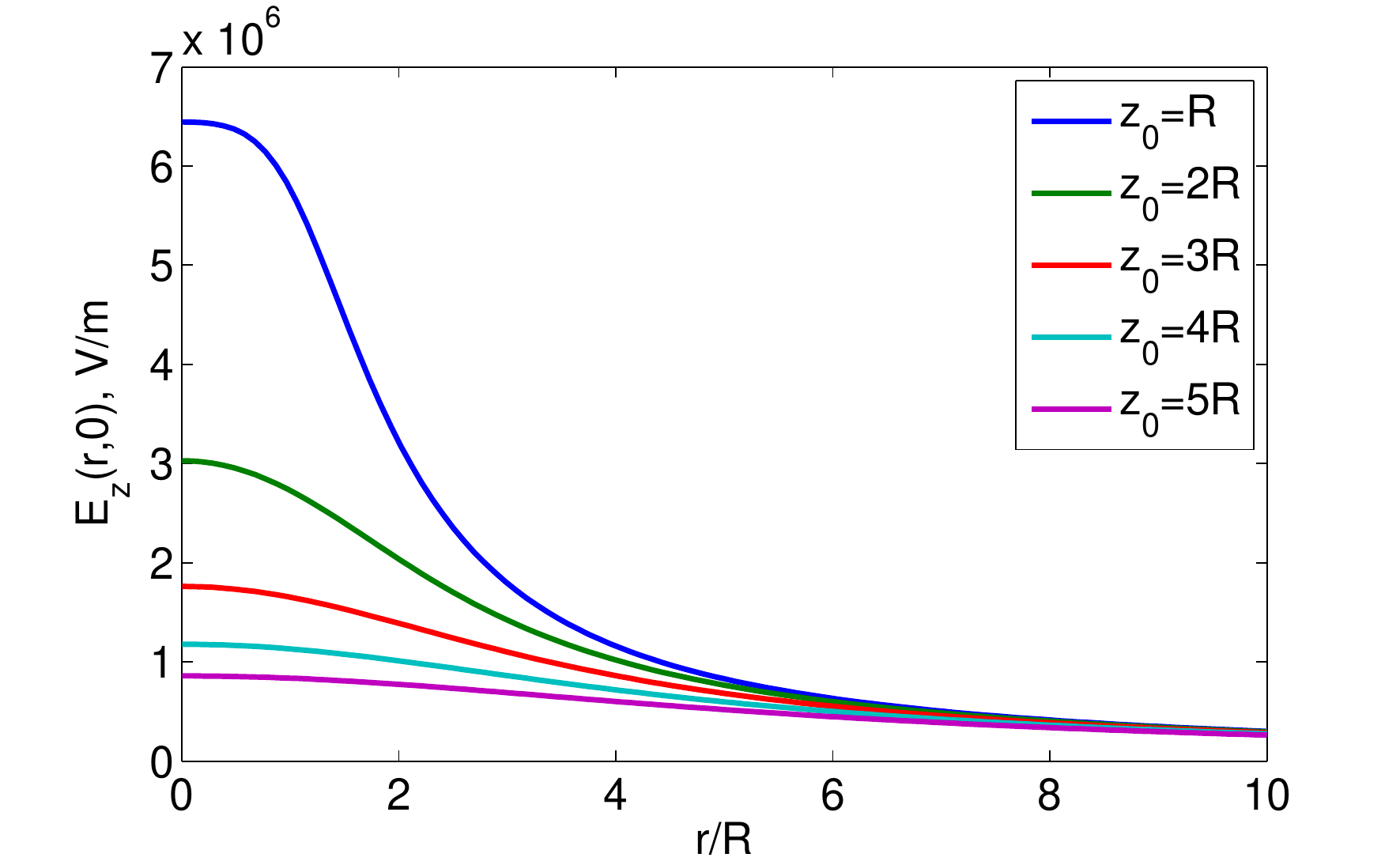}
  \includegraphics[width=3in]{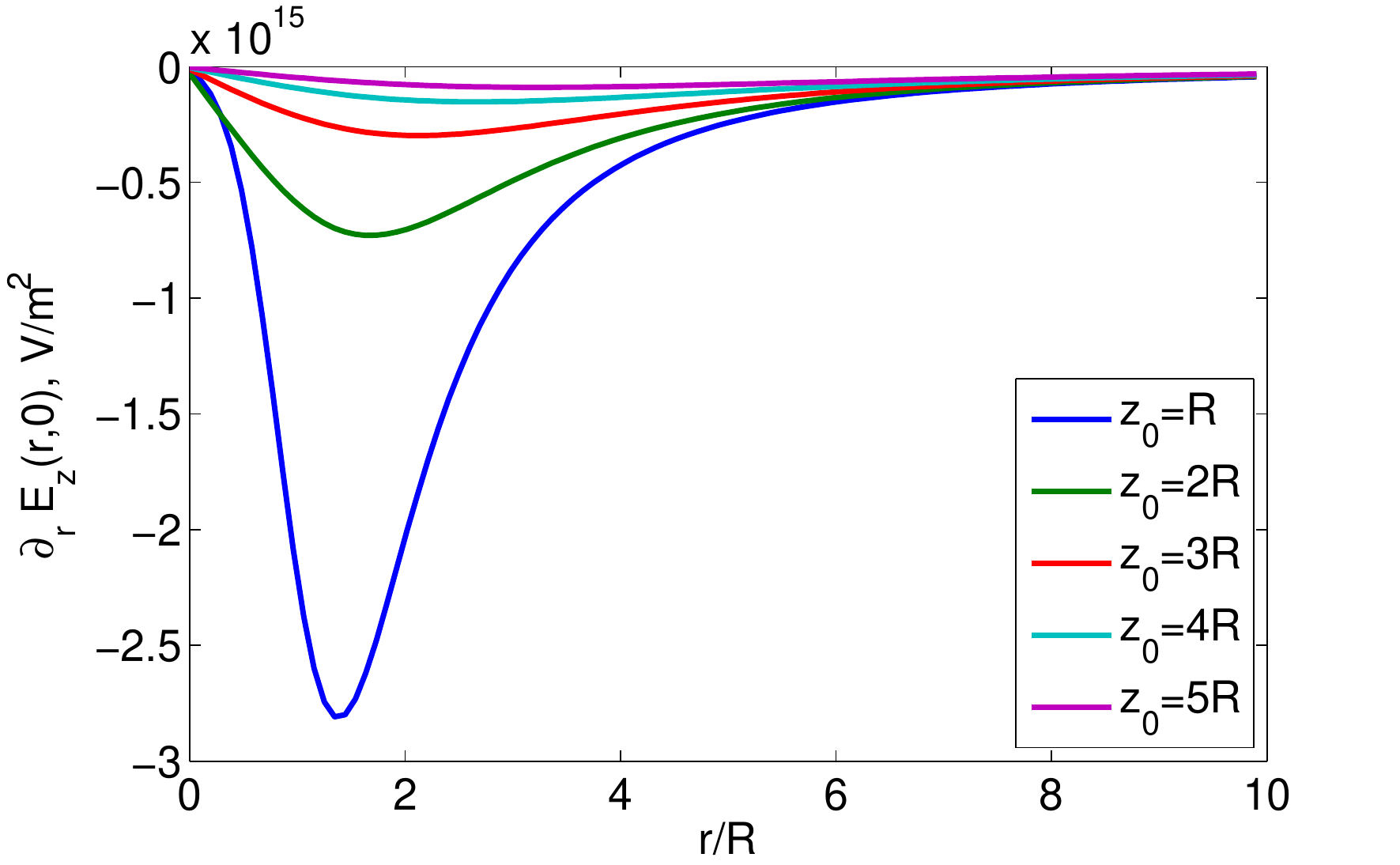}\\
  \caption{Electric field near the surface (left) and its radial gradient (right) as functions of the radial distance for different distances between the nanotube and the substrate.}\label{fig:field_surf}
\end{figure}

The magnitude of the electric field at the surface, obtained by numerically solving Eq. (\ref{eq:inteq}), is shown in Fig. \ref{fig:field_surf} (left panel) as a function of the radial distance, $r$, from the nanotube axis, and for different different values of the distance between the tip of the nanotube and the surface, $z_0$. We used the following parameters: $V=10$mV, $R=1$nm, and $L=50R$.
The value of the nanotube length that we used was found to be bigger than the intra-nanotube screening length, i.e. the results are valid also for much longer nanotubes.

The corresponding radial gradient of the electric field is shown in the right panel of Fig. \ref{fig:field_surf} and agrees by the order of magnitude with the simple estimate provided in Ref. \onlinecite{Puller}.
The coupling constant is obtain by multiplying this gradient by a typical molecular Stark shift (e.g.
$1$MHz per every kV/m for  dibenzanthanthrene (DBATT) molecules in hexadecane (HD)\cite{Brunel}.)

\section{Dissipation}
We now estimate the dissipation due to the Joule's heat originating from the currents induced in the the substrate and the nanotube when the latter vibrates.
For simplicity we assume that all of the nanotube charge is accumulated in its tip, in the region with length of the order of the nanotube radius.
This charge then can be estimated as\cite{LL}:
$Q(t)\approx\lambda(0,t)R\approx  VR/\log\left[4z(t)/R\right]$, where $z(t)$ is the distance between the tip and the substrate.

We model the displacement of the nanotube tip as that of a straight rod in $xz$ plane: $x(t)=x_{max}\cos\left(\Omega t\right)$, $z(t)=z_0+L-\sqrt{L^2-x^2(t)}\approx z_0+x^2(t)/(2L)$.
(We cannot use here the Euler-Bernoulli beam model,\cite{Cleland_book} since it ignores the displacement along the nanotube, i.e. effectively elongates it.)

{\em Dissipation by the intra-nanotube currents:}
The instantaneous power for the Joule heat generated by the currents flowing to and out of the nanotube tip is  $W(t)=\rho_{NT}L[\dot{Q}(t)]^2$, whereas the average energy dissipated during the period of the oscillations is
\bq \overline{W}_{tube}=\frac{\rho_{NT}}{8L}\left[\frac{Q_0}{z_0\log\left(\frac{4z_0}{R}\right)}\right]^2\om_M^2 x_{max}^4,\label{eq:dissipated_power}\eq
where $Q_0=VR/\log\left[4z_0/R\right]$, $\rho_{NT}$ is the nanotube resistance per unit length.

Note that the dissipated power is proportional to the forth power of the oscillations amplitude, which corresponds to a non-linear dissipative force $F_{Joul}=-m\gamma_{Joul}x^2\dot{x}$ (rather than the usual viscous friction force $F_{visc}=-m\gamma\dot{x}$).
This nonlinearity results from the geometric symmetry: the flow of charge in the nanotube is due to the change of the distance between the nanotube and the substrate, $z(t)-z_0$, which is proportional to second power of the oscillator displacement, $x(t)$.
A linear term would appear, if the equilibrium position of the nanotube was not perfectly vertical.

{\em Dissipation in the substrate:}
The nanotube oscillations induce redistribution of screening charge in the substrate, i.e. a flow of currents, which also dissipate energy via Joule heat.
For the typical nanotube oscillator frequencies (smaller than $1$GHz) the motion of the nanotube and screening charges is still sufficiently slow to be described as a quasi-electrostatic problem.
The Poisson equation now contains time-dependent charge density and potential, and needs to be supplemented by the appropriate time-dependent boundary conditions (relating for the instantaneous electric field outside and the inside the substrate to the surface charge), the current continuity equation, and the consistency relation between the current density and the electric field: $\mathbf{j}=\s_{sub}\mathbf{E}$ ($\s_{sub}$ is the substrate conductivity.)

The resulting solution can be shown to be similar to the one obtained by the method of images, but with the retarded image potential.
The dissipated energy is then obtained by integrating the local density of the Joule power, $w=\vec{j}\vec{E}=\s_c\vec{E}^2$, over the volume of the substrate.
The energy dissipated in the substrate per a period of the nanotube oscillations is then
\bq \overline{W}_{substrate}=\frac{\pi\s_{sub} Q_0^2 x_{max}^2}{3z_0^3}\frac{\om_M^2}{\om_M^2+(2\pi\s_{sub})^2}.\label{eq:diss_substrate}\eq

\begin{figure}[tbp]
  \includegraphics[width=3in]{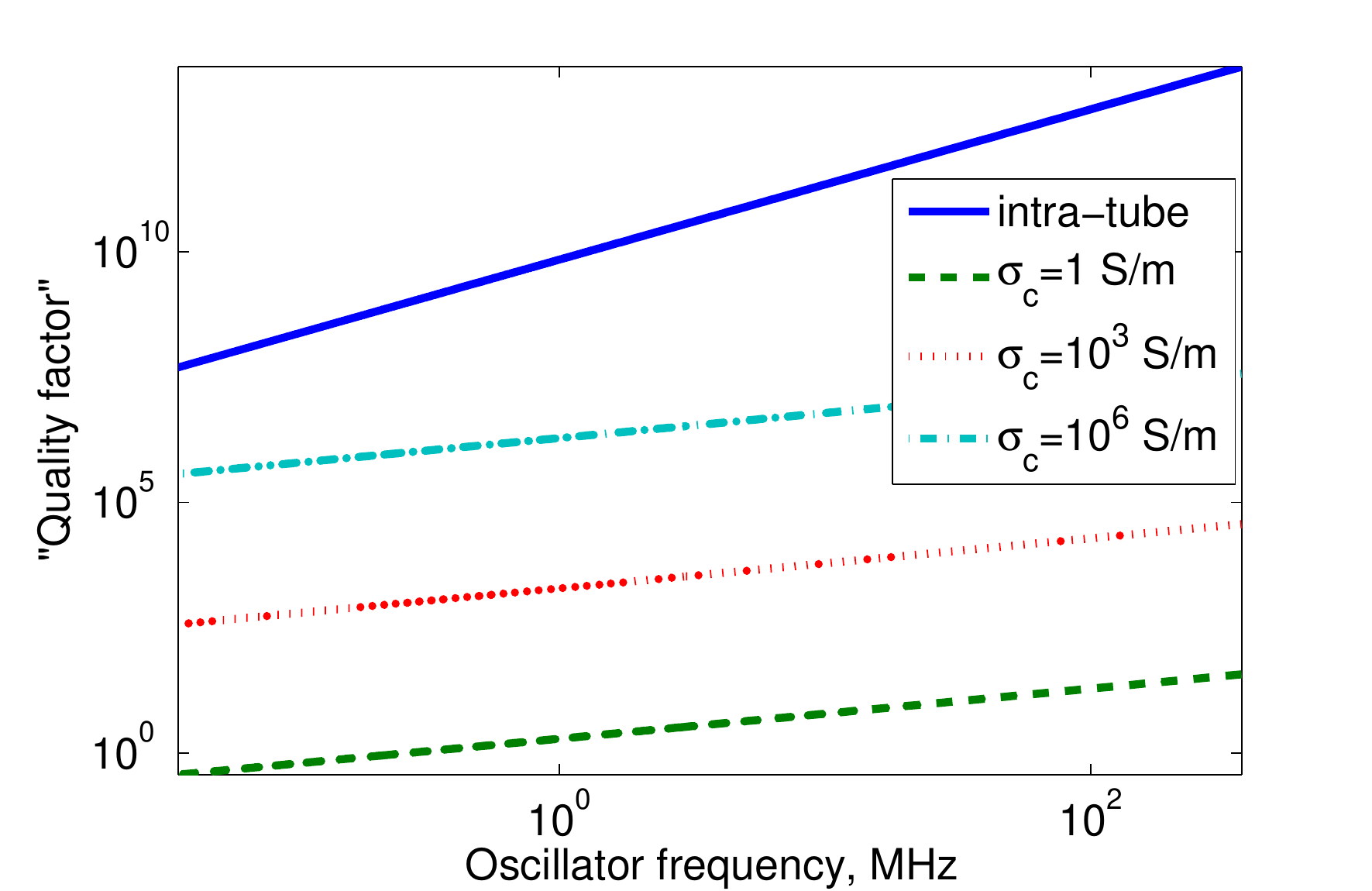}\\
  \caption{Fraction of the oscillator energy lost during one oscillation period due to the intra substrate currents (for $\s_c=1, 10^3, 10^6 S/m$) and due to the currents generated within the nanotube.}\label{fig:dissipation}
\end{figure}

The amplitude of thermal nanotube oscillations can be determined from the equipartition theorem:
$x_{max}^2\rightarrow \langle x^2\rangle=k_BT_K/(m\om_M^2)$.
The frequency of the main nanotube mode can be taken from the Euler-Bernoulli theory\cite{Poncharal,Purcell,Biedermann,Cleland_book}: $f_M=\om_M/(2\pi)=\mu_1^2\sqrt{\frac{E}{\rho}\left(d_o^2+d_i^2\right)}/(8\pi L^2)$,
where $\mu_1=1.875$, $E$ is the nanotube Young modulus,
$\rho=2300$kg/m$^{3}$
is the density of graphite, $d_o,d_i$ are the outer and the inner diameters of the nanotube.
We then obtain the estimates of the dissipation due to the intra-nanotube and substrate currents shown in Fig. \ref{fig:dissipation}, where we took $R=1$nm, $E=0.4$ TPa \cite{Poncharal}, $\rho_{NT}=6$MOm/m, and $T_K=1$K.
The quantity shown in Fig. \ref{fig:dissipation} is the ratio of the oscillator energy to the energy lost per one period of the oscillations due to the given dissipation mechanism, i.e. the effective quality factor.
We observe that the dissipation due to the substrate current, shown for several values of the substrate conductivity, is the dominant of the two dissipation mechanisms.
It can be quantified in a more practical way by defining the damping rate associated with energy loss of Eq. (\ref{eq:diss_substrate}):
\bq \gamma_{sub}= \frac{Q_0^2}{12\pi m z_0^3\s_{sub}}.\label{eq:gamma_sub}\eq
This equation is valid in the gaussian as well as in the SI units, and we took account that the conductivity of $1$S/m corresponds to frequency of $9$GHz, much greater than the possible frequency of the nanotube oscillator.
For the typical values of the parameters this equation gives the estimate $\gamma_{sub}=1$GHz nm$^{3}$ S/m, which means, e.g., that for $f_M=1$MHz oscillator removed by $z_0=10$nm from the substrate surface we need the conductivity of at least $\s_{sub}=10^3$S/m to ensure the quality factor of $Q=10^3$.

\section{Weak coupling, strong cooling, and weak back-action}
To order $\Omega_L^2$ the fluctuation spectrum for the population of the excited TLS state is given by
\bq S_{nn}(\om)\approx\int dte^{i\om t}\left\langle \tilde{n}(t)\tilde{n}(0)\right\rangle=
\frac{\Gamma\langle n\rangle}{\left(\delta-\om\right)^2+\frac{\Gamma^2}{4}},\eq
where $\tilde{n}=n-\langle n\rangle$.
Although it is possible to obtain the expression for this spectrum applicable to all orders in $\Omega_L$, it is rather cumbersome and would complicate comparison with the case when the oscillator is cooled by coupling to a photon cavity.\cite{Clerk_cooling,cooling}
However, one should keep in mind that the average population of the excited state of the molecule saturates at high laser strengths:
\bq\langle n\rangle=\frac{\Omega_L^2}{\delta^2+2\Omega_L^2+\Gamma^2/4}.\eq

The effective oscillator temperature is given by the following expression\cite{linearbackaction}
\bq \coth\left(\frac{\hbar\om_M}{2k_BT_{osc}}\right)=
\frac{\gamma\coth\left(\frac{\hbar\om_M}{2k_B T}\right)+\gamma_1\coth\left(\frac{\hbar\om_M}{2k_B T_{TLS}}\right)}
{\gamma+\gamma_1},\label{eq:cothTosc}\eq
where the cooling rate is determined by
\bq\gamma_1=\Delta_{zpf}^2
\left[S_{nn}(\om_M)-S_{nn}(-\om_M)\right]=
\Delta_{zpf}^2\langle n\rangle\Gamma
\left[\frac{1}{(\delta-\om_M)^2+\frac{\Gamma^2}{4}}-
\frac{1}{(\delta-\om_M)^2+\frac{\Gamma^2}{4}}\right],\label{eq:gamma1}\eq
whereas the effective TLS temperature is defined by
\bq \gamma_1 \coth\left(\frac{\hbar\om_M}{2k_B T_{TLS}}\right)=
\left[S_{nn}(\om_M)+S_{nn}(-\om_M)\right]=
\Delta_{zpf}^2\langle n\rangle\Gamma
\left[\frac{1}{(\delta-\om_M)^2+\frac{\Gamma^2}{4}}+
\frac{1}{(\delta-\om_M)^2+\frac{\Gamma^2}{4}}\right].\label{eq:gamma1coth}\eq
In the above expressions we have defined the coupling parameter $\Delta_{zpf}=\alpha x_{zpf}$,
where $x_{zpf}=\sqrt{\hbar/(2m\om_M)}$ is the zero-point fluctuation length for a quantum oscillator of frequency $\om_M$.

We now specify three conditions: the (i) {\em weak coupling} condition, $\gamma_1\ll\Gamma$, which specifies the limit where the effective temperature description by Eq. (\ref{eq:cothTosc}) is applicable\cite{Clerk_cooling}; (ii) the condition of absence of cooling, $\gamma_1\ll\gamma$; and (iii) the condition that one can neglect the effect of TLS fluctuations on the oscillator, $\gamma_1 T_{TLS}\ll\gamma T$ (which is the high-temperature limit of $\gamma_1 \coth\left[\hbar\om_M/(2k_B T_{TLS})\right]\ll\gamma \coth\left[\hbar\om_M/(2k_B T)\right]$.)

In the resolved side-band limit, $\om_M\gg\Gamma$, the first term in each of Eqs. (\ref{eq:gamma1}), (\ref{eq:gamma1coth}) takes maximum value at $\delta=\om_M$, whereas the other term is negligible and one arrives to the following conditions
\bsub\bn \rm{(i) }&& \Delta_{zpf}\ll \frac{\Gamma}{\sqrt{\langle n\rangle}},\\
\rm{(ii) }&&
\Delta_{zpf}\ll \sqrt{\frac{\gamma\Gamma}{\langle n\rangle}},\\
\rm{(iii) }&&
\Delta_{zpf}\ll
\sqrt{\frac{\gamma\Gamma}{\langle n\rangle}}
\frac{\sqrt{\langle x^2\rangle}}{x_{zpf}}=
\sqrt{\frac{\gamma\Gamma}{\langle n\rangle}}
\sqrt{\frac{2k_B T}{\hbar \om_M}}.\label{eq:resolvedsideband}\en\esub
Note that, although the excited state population $\langle n\rangle$ is also a function of the detuning, we treat it as an independent variable, since it can be adjusted by changing the laser strength.
Yet, we stress that $\langle n\rangle$ cannot exceed one half, which means that (unlike in the case of cooling with a photon cavity) one cannot at will change the validity of conditions Eq. (\ref{eq:resolvedsideband}).

For the parameters used throughout our manuscript and for $\om_M=\Gamma=10^8$Hz we have $x_{zpf}\approx 20$ pm (which is about two orders of magnitude smaller than $\sqrt{\langle x^2\rangle}\approx 1$nm.)
With the best possible value of $\alpha=10^{19}$Hz/m this gives us $\Delta_{zpf}\approx 10^8$ Hz.
Given that $\langle n\rangle\ll 1$ this means that we are always safely in the regime of weak coupling.
However, as we see from the second condition, {\em weak coupling does not exclude  strong cooling}, i.e. the regime in which $\sqrt{\gamma\Gamma/\langle n\rangle}\ll\Delta_{zpf}\ll \Gamma/\sqrt{\langle n\rangle}$.
Thus, the proposed setup can be used for cooling.

In absence of cooling, i.e. when $\alpha x_{zpf}\ll\sqrt{\gamma\Gamma/\langle n\rangle}$, the third condition is automatically satisfied, since $\sqrt{\langle x^2\rangle}/x_{zpf}\approx 50$, which means that one can neglect the TLS back-action on the oscillator for values of about $\alpha=10^{17}$Hz/m and less.

In the limit where the side-bands are not resolved, i.e. $\om_M\ll\Gamma$, maximizing $\gamma_1$ and $\gamma_1 T_{TLS}$ in respect to $\delta$ results in conditions
\bsub\bn \rm{(i) }&&
\Delta_{\Gamma}\ll \frac{\Gamma}{\sqrt{\langle n\rangle}},\\
\textrm{(ii) }&&
\Delta_{\Gamma}\ll \sqrt{\frac{\gamma\Gamma}{\langle n\rangle}},\\
\textrm{(iii) }&&
\Delta_{\Gamma}\ll \sqrt{\frac{\gamma\Gamma}{\langle n\rangle}}
\sqrt{\frac{2k_B T}{\hbar\Gamma}},\label{eq:unresolvedsideband}\en\esub
which differ from Eqs. (\ref{eq:resolvedsideband}) by the replacement
$x_{zpf}\rightarrow x_\Gamma=\sqrt{\hbar/(2m\Gamma)}$, $\Delta_{zpf}\rightarrow\Delta_\Gamma$.
Since in the previous estimate we took $\om_M=\Gamma$, we come to the same conclusions: one can neglect by the TLS back-action up to $\alpha=10^{17}$Hz/m, whereas for higher values one realizes regime of strong cooling, which however can be studied within the weak coupling approach employed here.

%\section{Specifying the limit of weak back-action}